\documentclass[aps,prd,reprint,longbibliography,nofootinbib,floatfix,superscriptaddress,author-year]{revtex4-2}  

\usepackage[portuges, english]{babel}
\usepackage[utf8]{inputenc}
\usepackage{amsmath}
\usepackage[makeroom]{cancel}
\usepackage{slashed}
\usepackage{enumitem}
\usepackage{times}
\usepackage{natbib}
\usepackage{xcolor}
\PassOptionsToPackage{hyphens}{url}
\usepackage[colorlinks=true,linkcolor=blue,citecolor=magenta,urlcolor=blue]{hyperref}
\usepackage{tikz}
\definecolor{lime}{HTML}{A6CE39}
\definecolor{Emerald}{HTML}{50c878}
\definecolor{PineGreen}{HTML}{01796F}
\definecolor{ForestGreen}{HTML}{228B22}
\definecolor{Coral}{HTML}{FF7F50}
\definecolor{YellowOrange}{HTML}{E94E16}  
\DeclareRobustCommand{\orcidicon}{%
	\begin{tikzpicture}
		\draw[lime, fill=lime] (0,0) 
		circle [radius=0.16] 
		node[white] {{\fontfamily{qag}\selectfont \tiny ID}};
		\draw[white, fill=white] (-0.0625,0.095) 
		circle [radius=0.007];
	\end{tikzpicture}
	\hspace{-2mm}
}
\foreach \x in {A, ..., Z}{%
	\expandafter\xdef\csname orcid\x\endcsname{\noexpand\href{https://orcid.org/\csname orcidauthor\x\endcsname}{\noexpand\orcidicon}}
}

\begin{document}
		
\medskip\noindent
\title{\large  
Constraining Ultra-Light Vector Bosons via Solar Neutrino Oscillations: \\ The Gauged $L_\mu - L_\tau$ Model and Prospects for JUNO and XLZD}
		
		\author{Ilídio Lopes\orcidA{}}
		\affiliation{Centro de Astrof\'{\i}sica e Gravita\c c\~ao  - CENTRA, \\
			Departamento de F\'{\i}sica, Instituto Superior T\'ecnico - IST,\\
			Universidade de Lisboa - UL, Av. Rovisco Pais 1, 1049-001 Lisboa, Portugal}
		\email{ilidio.lopes@tecnico.ulisboa.pt}

\begin{abstract}
\noindent		
We present an extension of the Standard Model based upon the gauged $U(1)_{L_\mu - L_\tau}$ symmetry, featuring an ultra-light vector boson that couples to the difference of muon and tau lepton numbers. This anomaly-free construction, whilst historically motivated by the muon's anomalous magnetic moment, is here employed as an independent phenomenological framework for solar neutrino physics. Through kinetic mixing with the photon, the $Z^\prime$ boson generates an effective matter potential within the solar interior, thereby modifying neutrino flavour oscillations via the Mikheyev--Smirnov--Wolfenstein mechanism.
	
\noindent	
Utilising recent solar neutrino measurements from Borexino, SNO, and Super-Kamiokande, together with an up-to-date solar standard model, we constrain the boson mass to be below $5.7\times 10^{-17}\,{\rm eV}$, positioning it as a potential ultra-light dark matter candidate. The optimal model, corresponding to a coupling parameter $\eta_o \approx -2$, yields $\chi_\nu^2 = 2.7$, which compares favourably with the standard three-flavour scenario ($\chi_\nu^2 = 3.1$). This best-fit configuration corresponds to an effective four-fermion coupling $g_{Z^\prime}^2 \varepsilon / m_{Z^\prime}^2 \approx 6.6 \times 10^{-23}\,{\rm eV}^{-2}$, approximately $5.7$ times the Fermi constant, which for $\varepsilon \approx 1$ and a reference mediator mass of $1\,{\rm eV}$ is equivalent to a gauge coupling $g_{Z^\prime} \approx 8\times 10^{-12}$.
	
\noindent	
We furthermore present projections for forthcoming experiments. The Jiangmen Underground Neutrino Observatory (JUNO), which commenced operations in August 2025 and has already achieved world-leading precision in oscillation parameter measurements, together with the XLZD consortium, should substantially refine these constraints. Such ultra-light particles may address long-standing discrepancies in galaxy simulations within the cold dark matter paradigm, and recent theoretical developments demonstrate their compatibility with non-thermal dark matter production mechanisms. Additionally, terrestrial long-baseline neutrino experiments provide complementary probes of this parameter space.
\end{abstract}


\maketitle
		
\section{Introduction}
\label{sec:intro}
	\noindent	
	In recent years, our comprehension of the Universe has been subjected to rigorous scrutiny, driven by breakthroughs in particle physics experiments alongside cosmological and astrophysical observations. Within this context, the standard model of elementary particles and fundamental interactions finds itself under considerable examination, motivating physicists to explore theoretical frameworks that extend beyond the conventional paradigm. 
			
			\medskip\noindent
			The dark matter problem remains amongst the most compelling motivations for physics beyond the Standard Model \citep{2013PhR...531....1S,2018PhR...761....1B,2021PrPNP.11903865A,2021ScPP...10...30B,2015ARNPS..65..485G}. Despite decades of experimental effort, the fundamental nature of dark matter remains elusive, and its existence is inferred solely through gravitational effects on cosmological and galactic scales. Ultra-light bosonic particles represent a particularly intriguing class of dark matter candidates, capable of addressing several outstanding puzzles in structure formation \citep{2021ARA&A..59..247H}.
			
			\medskip\noindent
			The present study focuses on neutrinos, which occupy a rather singular position amongst fundamental particles owing to their distinctive behaviour and the many unresolved mysteries they present \citep[see, e.g.,][]{1987RvMP...59..671B,2004NJPh....6..122M,2022PrPNP.12403947A}. It is precisely for this reason that a new generation of diverse neutrino experiments is now underway \citep{2022PrPNP.12403947A,2023PrPNP.13104043X}, with the aim of determining neutrino parameters and characteristics to unprecedented accuracy, particularly those pertaining to the three-neutrino flavour oscillation model. This unique experimental landscape positions neutrino physics as an invaluable instrument for exploring new particles and interactions beyond the standard model.

			\medskip\noindent	
			In light of these experimental advancements, the scientific community has been motivated to formulate a considerable variety of neutrino models \citep[see, e.g.,][]{1991PhLB..271..172G,2000PhRvD..62g3001B,2002NuPhB.629..479G,2004PhRvD..70c3010G,2015NJPh...17i5002M,2018FrP.....6...10T,2023JHEP...07..071A}, commonly referred to as Non-Standard Interaction (NSI) models. These theoretical constructions are poised for rigorous evaluation by the forthcoming generation of neutrino detectors \citep{2022PrPNP.12403947A,2023PrPNP.13104043X}.
			
			\medskip\noindent	
			This article presents a Non-Standard Interaction model based on the gauged $U(1)_{L_\mu - L_\tau}$ symmetry, featuring a vector boson whose mass may range from ultra-light to massive. Unlike models that rely on complex baryon-number-dependent charge assignments, the $L_\mu - L_\tau$ gauge symmetry is naturally anomaly-free without requiring new fermions or ad-hoc charge assignments \citep{2011JPhG...38h5005H}. The charge assignments under this symmetry are elegantly simple: the muon sector ($\nu_\mu, \mu$) carries charge $+1$, the tau sector ($\nu_\tau, \tau$) carries charge $-1$, whilst the electron sector and all quarks remain uncharged. This model was historically motivated by hints of lepton-flavour-universality violation in $B$-meson decays and the muon's anomalous magnetic moment. It is worth noting, however, that the experimental and theoretical landscape has evolved considerably since these anomalies were first identified. The LHCb collaboration's improved measurements of $R_{K}$ and $R_{K^{\!*}}$, announced in December 2022, are now consistent with the Standard Model within $1\sigma$ \citep{2023EPJC...83..648A}. Similarly, the Fermilab Muon g-2 experiment's final measurement, combined with advances in lattice QCD calculations, has essentially eliminated the previously reported discrepancy between theory and experiment \citep{2025PhR..1143....1A}. Consequently, whilst these historical anomalies provided the original impetus for the model, its current utility lies in its unique ability to explain dark matter phenomenology and potential long-range forces in the neutrino sector. Our present analysis examines its application to neutrino-plasma interactions in the Sun's interior, thereby establishing an independent phenomenological testing ground.
			
			\medskip\noindent	 
			Through our investigation, we determine the ideal mass range for these bosons that ensures alignment with existing solar neutrino experimental measurements \citep[see, e.g.,][]{2017PhR...685....1W,2022PrPNP.12403947A}. Our analysis identifies the boson as an ultra-light particle. We furthermore project the potential constraints on this model by drawing upon the expected accuracy of forthcoming solar neutrino experiments \citep[e.g.,][]{2021ARNPS..71..491O}, most notably those from the DARWIN/XLZD \citep{2024NuPhB100316473B,2020EPJC...80.1133D} and JUNO \citep{2020arXiv200611760J, 2022arXiv221008437J,2023arXiv230303910A} collaborations.
			
\medskip\noindent	
The identification of the boson as an ultra-light particle constitutes a significant finding, one that offers potential solutions to several outstanding astrophysical and cosmological problems. Such a boson presents itself as a promising dark matter candidate \citep{2021ARA&A..59..247H}, with implications for the cosmic evolution of the Universe. Additionally, it may contribute towards resolving the solar abundance problem \citep[see, e.g.,][]{2008PhR...457..217B,2016EPJA...52...78S}, a persistent inconsistency amongst solar models, observed neutrino fluxes, and helioseismic observations that has resisted resolution despite considerable effort over the past two decades. Furthermore, recent phenomenological developments have demonstrated that the $L_\mu - L_\tau$ model can accommodate non-thermal dark matter production and leptogenesis \citep{2025PhRvD.111c5008Q}, potentially explaining the baryon asymmetry of the Universe. 
		
 \begin{figure*} 
 	\includegraphics[scale=0.5]{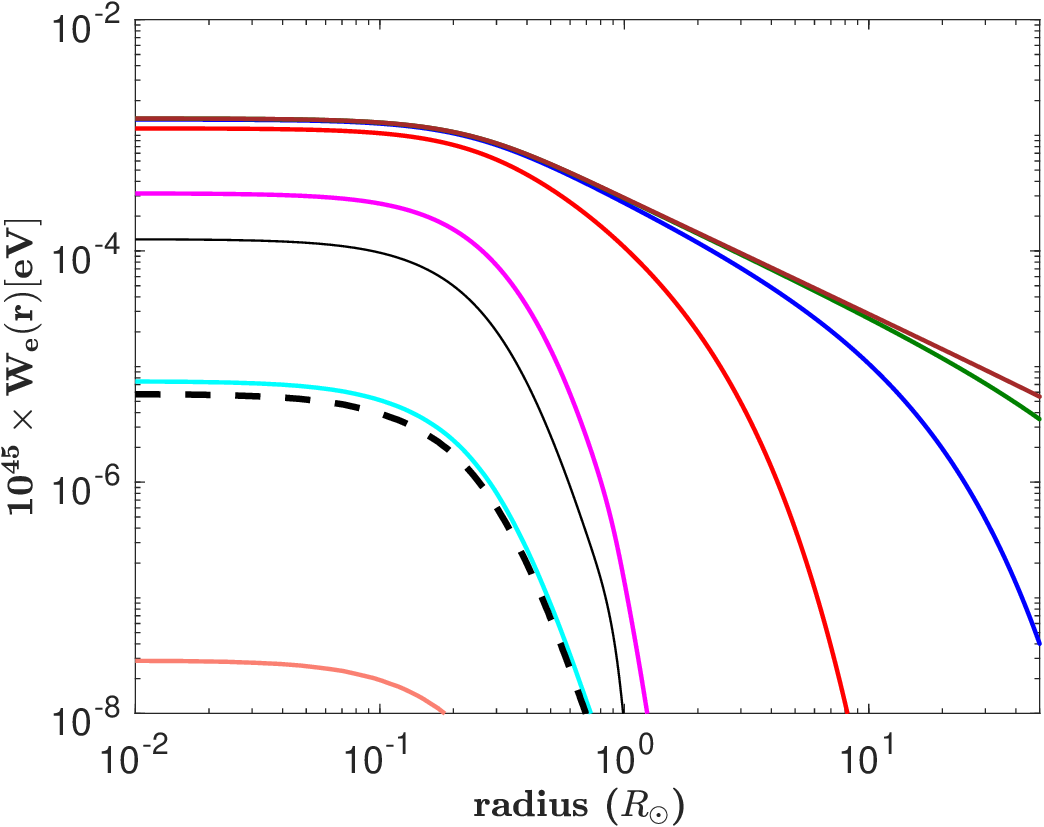}	
 	\hspace{0.1cm}
 	\includegraphics[scale=0.5]{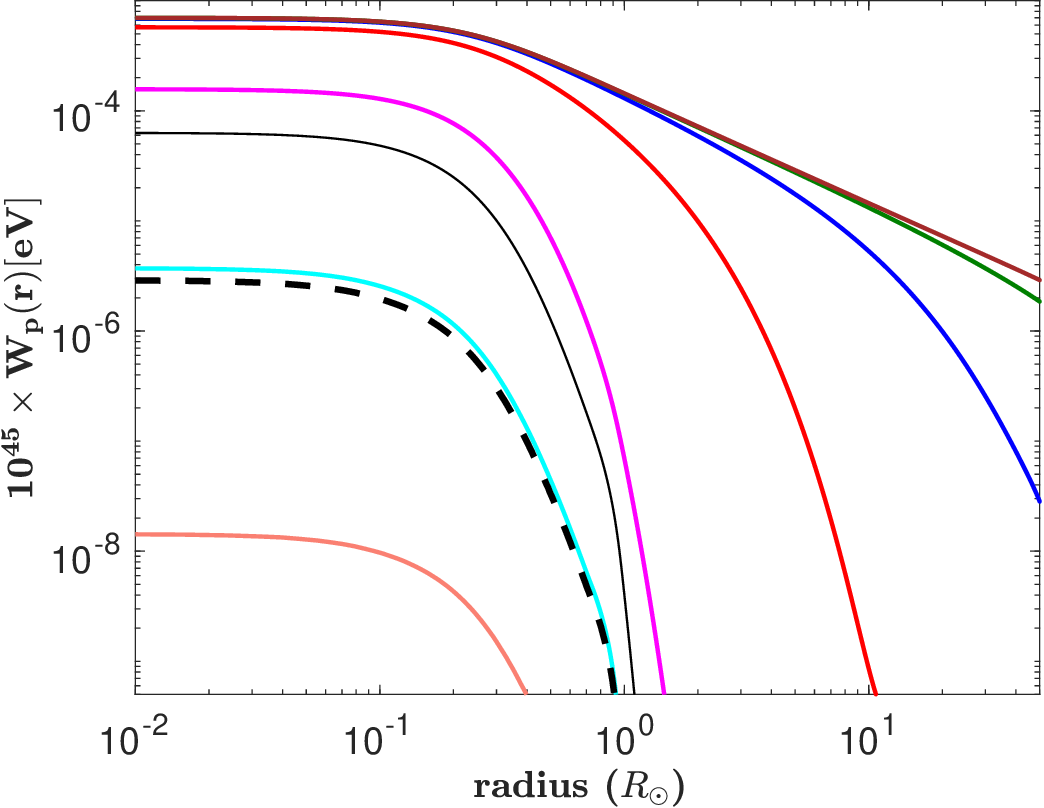}	
 	\caption{The electron (left panel) and proton (right panel) potential functions, $W_e(r)$ and $W_p(r)$, inside the Sun as a function of distance from the solar centre, expressed in units of the solar radius $R_\odot$. The curves correspond to a range of $\lambda_{Z^\prime}/R_\odot$ values, with colours transitioning from salmon (lowest curves) to brown (uppermost curves); these correspond to $\lambda_{Z^\prime}/R_\odot = 10^{-3}$, $10^{-2}$, $10^{-1}$, $1$, $10$, $10^2$, and $10^3$, respectively. The dashed black line serves as a reference, representing the scaled internal electron density ($10^{-3}\,n_e$) in the left panel and proton density ($10^{-3}\,n_p$) in the right panel, as obtained from the standard solar model.}
 	\label{fig:LambdaBoson}
 \end{figure*}

\section{Model Overview}
\label{sec:Model}

\noindent	
Here, we expand upon the Lagrangian of the standard model by introducing the gauged $U(1)_{L_\mu - L_\tau}$ symmetry, accompanied by a lightweight gauge boson $Z^\prime$ that couples to the difference of muon and tau lepton numbers. Since the first generation of leptons and all quarks have zero charge under this symmetry, the $Z^\prime$ boson does not directly couple to the dominant matter in the Sun. To generate a matter potential relevant for solar neutrinos, this model relies on \emph{kinetic mixing} between the $Z^\prime$ field tensor and the electromagnetic field tensor. The interaction Lagrangian is given by \citep{2011JPhG...38h5005H}:
\begin{eqnarray}
	{\cal L}_{Z^\prime}	= g_{Z^\prime} Z^\prime_{\alpha} \left(\bar{\mu}\gamma^\alpha \mu - \bar{\tau}\gamma^\alpha \tau + \bar{\nu}_\mu \gamma^\alpha \nu_\mu - \bar{\nu}_\tau \gamma^\alpha \nu_\tau\right)
	\nonumber
	\\ + \frac{\varepsilon}{2} F^{\mu\nu} Z^\prime_{\mu\nu},
	\label{eq:LNSI2}
\end{eqnarray}
where $g_{Z^\prime}$ is the coupling constant and $\varepsilon$ is the kinetic mixing parameter.
Due to the kinetic mixing term, the $Z^\prime$ boson acquires a coupling to the electromagnetic current proportional to $\varepsilon e$. Consequently, the dense plasma of protons and electrons in the Sun generates a non-zero potential for the neutrinos. This potential distinguishes clearly between $\nu_\mu$ and $\nu_\tau$, generating a diagonal potential matrix ${\rm diag}(0, V, -V)$ in the flavour basis. For a light mediator, finite-size effects of the matter distribution are captured by the convolution $W_i(r)$ ($i = e, p$) in Eq.~(\ref{eq:nisphere}) \citep{2019JHEP...12..046S,2019arXiv190700991B}.

\noindent	
The $L_\mu - L_\tau$ gauge symmetry is naturally anomaly-free without requiring new fermions or ad-hoc baryonic charge assignments \citep{2011JPhG...38h5005H,2015PhRvD..91g5006C}. Our focus is on how this Non-Standard Interaction (NSI) shapes the neutrino flavour oscillation through the kinetic mixing portal. Consequently, neutrinos in this model interact coherently with the solar plasma \citep{2020PhLB..80335349F}. It is important to note that since the neutron content of the Sun does not contribute to the kinetic mixing potential, only the proton (and electron) density enters the effective potential calculation.

\medskip\noindent
In this study, we investigate neutrino propagation through both vacuum and matter, following the standard three-flavour neutrino oscillation model. This approach extends existing work, as referenced in various studies \citep[e.g.,][]{1989RvMP...61..937K, 2013JHEP...09..152G,2020ApJ...905...22L}. Our research builds upon the well-established Mikheyev-Smirnov-Wolfenstein effect (MSW) \citep{1978PhRvD..17.2369W,1985YaFiz..42.1441M}, whilst taking into consideration the presence of a $Z^\prime$ vector boson mediator as introduced by \citet{2023PhRvD.108c5028L}. In the $L_\mu - L_\tau$ model, neutrinos interact with the solar plasma through the kinetic mixing portal. Due to the kinetic mixing, the $Z^\prime$ boson couples to the electromagnetic current, and consequently the local potential of charged particles within the solar medium influences neutrino scattering, as characterised by the functions $W_i(\mathbf{r})$ with $i = e, p$ \citep{2007JCAP...01..005G,2019JHEP...12..046S}. Assuming spherical symmetry within the Sun, $W_i(\mathbf{r})$ simplifies to $W_i(r)$:
\begin{align}
	W_i(r) = \frac{2\pi \lambda_{Z^\prime}}{r}
	\int_0^{R_\odot} r^{\prime}
	n_{i}(r^\prime)
	\left[ e^{- \frac{|r^\prime-r|}{\lambda_{Z^\prime}} }
	- e^{-\frac{|r^\prime+r|}{\lambda_{Z^\prime}}} \right]
	\mathrm{d}r^{\prime},
	\label{eq:nisphere}
\end{align}  
\noindent 
where $i = e, p$ denotes electrons and protons respectively, and $\lambda_{Z^\prime}=1/m_{Z^\prime}$ delineates the range of the interaction. In the $L_\mu - L_\tau$ model with kinetic mixing, only the proton density $n_p(r)$ contributes to the effective potential, as the neutron content of the Sun does not couple to the electromagnetic current. Crucially, $W_p(r)$ captures the modulation of the matter potential attributable to the finite interaction range of the $Z^\prime$ vector boson, thereby differentiating from the conventional contact interaction. Complementary constraints on light-mediator NSI arise from neutrino scattering, in particular, coherent elastic neutrino-nucleus scattering, and can be combined with oscillation data \citep{2023JHEP...08..032C,2019arXiv190700991B}.
Figure~\ref{fig:LambdaBoson} shows the influence of the parameter $\lambda_{Z^\prime}$ on $W_e(r)$ and $W_p(r)$.

\medskip\noindent
Here, we express the influence of non-standard interactions (NSI) on neutrino propagation as a contribution to the potential, as follows:
\begin{eqnarray}
	V_{\rm eff}= V_e(r)-V_{p}(r)	
	=\sqrt{2}\,G_F\, n_{\rm eff}(r),
	\label{eq:Veff}
\end{eqnarray}
where $V_e=\sqrt{2}\,G_F\, n_e$ and $V_{p}=\sqrt{2}\,G_F\,  \zeta_{Z\prime}$.
In this context, $n_{\rm eff}\equiv n_e(r)-\zeta_{Z\prime}(r)$, $G_F$ is the Fermi constant, $n_e(r)$ is the radial profile of electron density, and $\zeta_{Z\prime}(r)$ represents the NSI strength term. In the $L_\mu - L_\tau$ model with kinetic mixing, only the proton density contributes:
\begin{eqnarray}
	\zeta_{Z\prime}(r)= \eta_{o} W_p(r).	
	\label{eq:zetaZ}
\end{eqnarray}
Here, $\eta_{o}= g_{Z^\prime}^2 \varepsilon/(2\sqrt{2}\, G_F\, m_{Z^\prime}^2)$ is a dimensionless parameter that captures the relationship between the gauge coupling $g_{Z^\prime}$, the kinetic mixing $\varepsilon$, and the mediator mass $m_{Z^\prime}$; here $W_p(r)$ denotes the effective (mediator-smeared) proton number density, so that $\zeta_{Z^\prime}(r)$ carries the dimensions of a number density. Conveniently, and following \citet{2007JCAP...01..005G}, the associated potential is $V_{p}= g^2_{Z^\prime}\varepsilon\, W_p(r)/(2\, m_{Z^\prime}^2)$, the factor of one half arising directly from the definition of $\eta_{o}$.

\begin{figure}[!t]
\centering 
\includegraphics[scale=0.5]{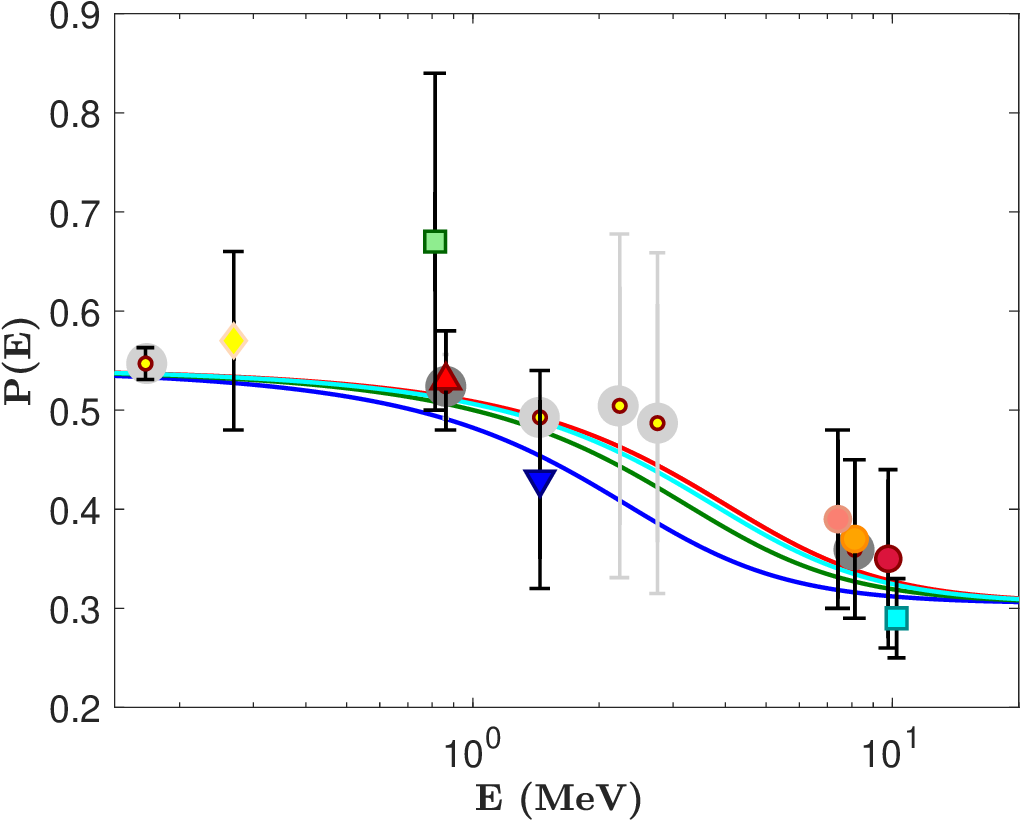}	
	\caption{The plot illustrates the survival probability of electron neutrinos for various models. The red curve represents the standard solar neutrino model ($\eta_o=0$). NSI models are depicted as follows: $\lambda_{Z^\prime}/R_\odot=1$ with $\eta_o=-6$ (blue curve) and $\eta_o=-2$ (green curve), and $\lambda_{Z^\prime}/R_\odot=0.1$ with $\eta_o=-2$ (cyan curve). The coloured markers signify the measured survival probabilities from three solar neutrino detectors (SNO, Super-Kamiokande, and Borexino) based on the prevailing standard solar model. The grey-yellow markers forecast the anticipated measurements from the upcoming DARWIN/XLZD and JUNO collaborations. Error bars for current and anticipated data reflect uncertainties inherent in the measurements and those arising from the current standard solar model. Further details and data source references are provided in the main text.
	}
	\label{fig:Pe}
\end{figure}

\medskip\noindent
We are examining the probability of electron neutrinos surviving in different models with varying values of $\eta_o$. We then compare these findings with the latest solar neutrino experiment data. If you are interested in more standard calculations, see the recent particle physics review \citep{2018PhRvD..98c0001T,2016ChPhC..40j0001P}, specifically the `Neutrino Masses, Mixing, and Oscillations' section from the August 2024 update \citep[][]{2024PhRvD.110c0001N}. In simple terms, we use a formula to predict how likely electron neutrinos are to survive: $P_{e}(E)\approx\cos^4{(\theta_{13})}P_{e}^{2\nu_e}+\sin^4{(\theta_{13})}$.
We calculate the survival probability of electron neutrinos in a two neutrino flavour model as $ P_{e}^{2\nu_e}(E)=0.5+\left(0.5-P_{\gamma}\right)\cos{(2\theta_{12})} \cos{(2\theta_{m})}$. Here, $P_{\gamma}$ accounts for probability jumps due to the non-adiabatic correction, and $\theta_{m}$ is the matter mixing angle \citep{2003NIMPA.503....4D}, calculated at the neutrino source region in the Sun
\citep[e.g.,][]{1989PhRvD..39.1930K,1995PhRvD..51.4028B}.
The matter mixing angle, $\theta_m$, is computed as $\cos(2\theta_{m})={A_m}/{({A_m^2 +\sin^2{(2\theta_{12})}})^{1/2}}$. $A_m$ depends on the initial mixing angle $\theta_{12}$, the neutrino energy $E$, and the effective matter potential $V_{\rm eff}$ (equation \ref{eq:Veff}), in combination with the squared-mass difference $\Delta m^2_{21}$. We compute $A_m$ as $A_m=\cos{(2\theta_{12})}-{2 E\, V_{\rm eff}}/{\Delta m^2_{21}}$, the energy factor rendering the second term dimensionless, as required. For a detailed discussion, refer to \citet{2023PhRvD.108c5028L}.

\medskip\noindent 
The probability of an electron neutrino surviving various nuclear reactions depends on where it originates within the Sun. For more on how the source location affects the survival probability $P_e(E)$, see \citet{2013PhRvD..88d5006L,2017PhRvD..95a5023L}. The mean survival probability for each nuclear reaction in the Sun's interior, denoted as $P_{e,k} $, is calculated as follows:
\begin{eqnarray}
	P_{e,k} (E) =A_k^{-1} \int_0^{R_\odot} P_{e} (E,r)\phi_k (r) 4\pi \rho(r) r^2 dr.
	\label{eq:Peek}
\end{eqnarray}
In this equation, $A_k$ serves as a normalisation constant.  
The variable $k$ denotes the solar neutrino sources, including $pp$, $pep$, $^8B$, $^7Be$, $^{13}N$ and $^{15}O$.
Several factors, such as vacuum and matter oscillations and the Sun's internal physics, affect whether electron-neutrinos change flavour. Local solar plasma conditions have a significant role in this flavour conversion.

\section{Present Constraint}
\label{sec:CurrentData}
		
\begin{figure} 
\centering 
\includegraphics[scale=0.5]{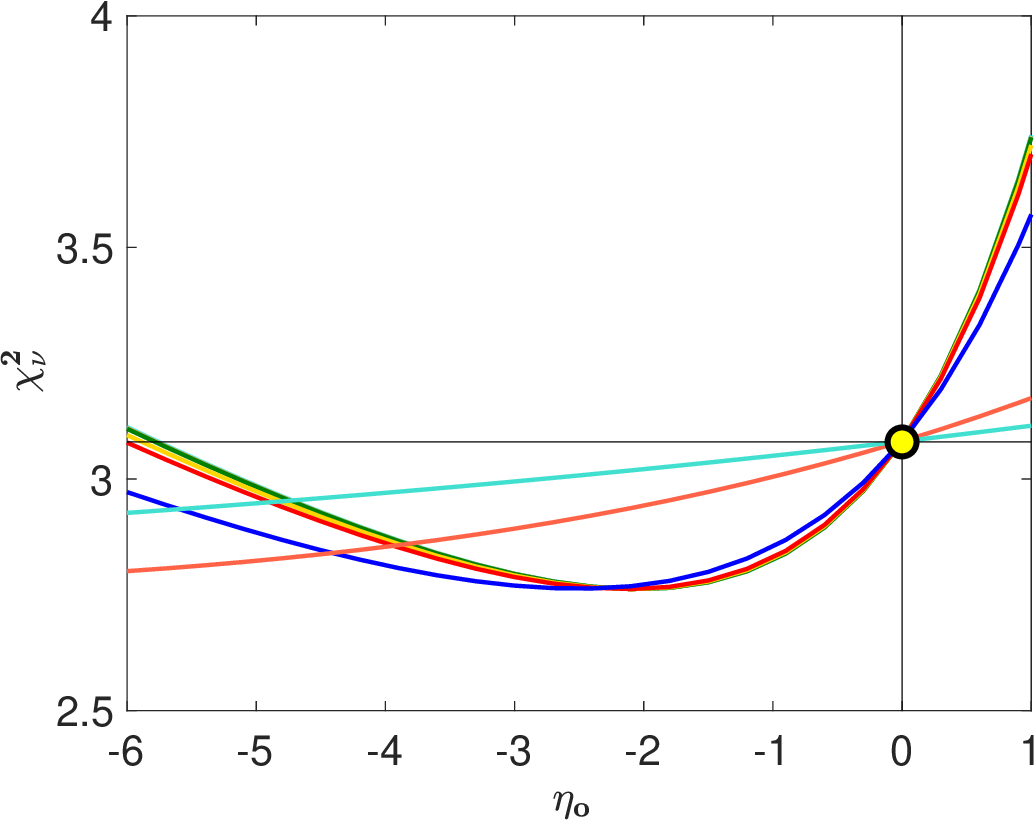}
\caption{Graph depicting the $\chi^2_\nu$ test values versus the coupling constant $\eta_o$ across various NSI neutrino models. Continuous lines indicate the $\chi^2_\nu$ values for distinct $\lambda_{Z^\prime}/R_\odot$ values, with colour coding as follows: 0.05 (Turquoise), 0.1 (Tomato), 1 (Blue), 5 (Red), 10 (Gold), 50 (Green), 100 (Light Brown), and 1000 (Cyan). The vertical line corresponding to $\eta_o=0$ represents the standard model, while the horizontal line at $\chi^2_\nu=3.1$ indicates the quality of fit for the current data. The yellow circle represents the $\chi^2_\nu$ of the current neutrino flavour oscillation model.}
\label{fig:NusundataChi2}
\end{figure}
		
		\medskip\noindent
		Figure \ref{fig:Pe} compares our predictions for the survival probability of electron neutrinos from several non-standard interaction (NSI) models to the current solar neutrino data. The data analysis was conducted using the High-Z solar standard model, which is a well-validated model of the Sun's interior \citep[e.g.,][and references therein]{2020MNRAS.498.1992C,2017ApJ...835..202V}. Each data point in the figure represents the observed survival probability of electron neutrinos as measured by three major solar neutrino detectors: SNO, Super-Kamiokande, and Borexino. The uncertainty associated with each data point is a combination of experimental error and theoretical uncertainty, which arises from modelling the Sun's interior \citep[e.g.,][and references therein]{2020MNRAS.498.1992C,2021ARNPS..71..491O}.
		More specifically, Borexino data includes measurements from various solar neutrino reactions, such as $pp$  (yellow diamond),  
		$^7Be$ (red upward-triangle),  $pep$ (blue downward-triangle), and  
		$^8B$ in the High-Energy Region (HER), denoted by circles in salmon (HER), orange (HER-I), and magenta (HER-II) colors; SNO's  
		$^8B$ measurements are indicated by a cyan square; and the joint KamLAND/SNO   $^7Be$  measurements are represented by a green square. Please refer to \citet{2018Natur.562..505B,2019PhRvD.100h2004A,2010PhRvD..82c3006B,2011PhRvC..84c5804A,2016PhRvD..94e2010A,2013PhRvC..88b5501A,2008PhRvD..78c2002C} and associated references for more in-depth understanding of this experimental data. For reference, the range of the MSW-LMA solution (within the $1-\sigma$ variation) is significantly smaller than the uncertainties associated with the current data points, as indicated by \citet{2017PhRvD..95i6014C}.
		
		\medskip\noindent
		To  strengthen our analysis, we introduce a test analogous to the chi-squared test, denoted as $\chi_\nu^2$. This test leverages the relationship of the electron neutrino survival probability with the solar background structure. The test is formulated as:
		\begin{equation}
			\chi^2_{\nu}= \sum_{i,k} \left(\frac{P_{e,k}^{obs}(E_{i})-P_{e,k}^{th}(E_{i})}{\sigma_{obs(E_{i})}} \right)^2,
			\label{eq:Chi2nu}
		\end{equation}
		where this metric juxtaposes our theoretical models against empirical data accumulated from diverse neutrino experiments. Within this framework, different energy levels, denoted by $E$, are used to ascertain the survival probability function $P_{e,k}(E)$, detailed further in equation (\ref{eq:Peek}). Here, 'obs' represents the observed readings, and 'th' stands for the theoretical predictions for the neutrino energy $E_i$. The indices $i$ and $k$ respectively signify particular experimental data points and the corresponding solar neutrino sources. The term $\sigma_{obs}(E_{i})$ designates the associated measurement error for the \(i^{th}\) entry. The values $P_{e,k}^{obs}(E_{i})$ emanate from empirical findings of solar neutrino experiments.
\begin{figure}
	\centering
	\includegraphics[scale=0.45]{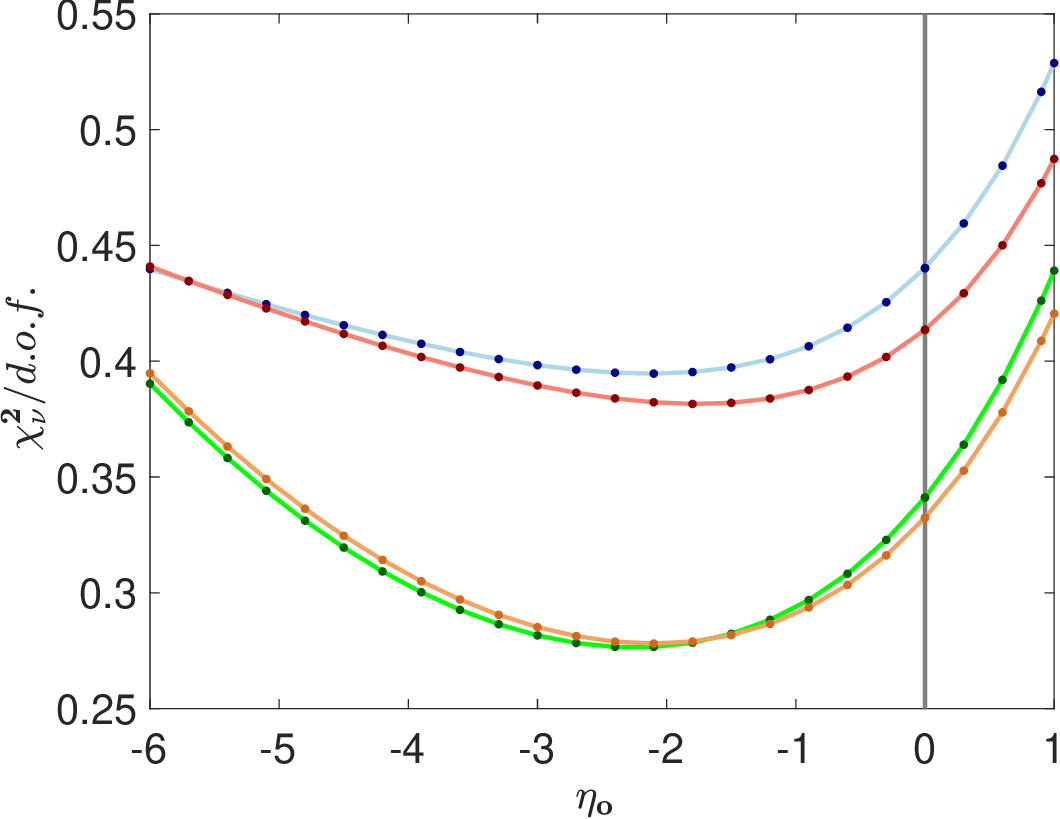} 
	\caption{Comparison of $\chi^2_\nu/\text{d.o.f.}$ across various NSI models for $\lambda_{Z^\prime}/R_{\odot}=5.0$ as a function of $\eta_o$. The curves represent the anticipated $\chi^2_\nu/\text{d.o.f.}$ precision based upon distinct data sets: (i)~current data (blue); (ii)~current data augmented with expected DARWIN/XLZD $pp$ measurements (red); (iii)~current data enriched with forthcoming JUNO measurements (green); and (iv)~a comprehensive set including current data, DARWIN/XLZD $pp$, and JUNO measurements (orange). A detailed discussion is provided in the main text.}
	\label{fig:Chi2All}
\end{figure}

\medskip\noindent 
Figure \ref{fig:NusundataChi2} showcases how $\chi_\nu^2$ varies for a selected group of NSI models with respect to $\eta_o$ and $\lambda_{Z^\prime}$. Based on equation \ref{eq:Chi2nu}, the standard neutrino flavour model (where $\eta_o=0$) gives a $\chi_\nu^2$ of $3.1$, which is denoted by a yellow circle in the figure.  The plotted curves correspond to diverse values of $\lambda_{Z^\prime}$ (or its inverse, $m_{Z^\prime}$), spanning from $0.05$ to $1000$. We discern two pronounced asymptotic trends: 
\begin{enumerate}
\item For diminished values of $\lambda_{Z^\prime}$ (or elevated values of $m_{Z^\prime}$), the influence recedes. As shown in equation \ref{eq:Veff}, the contribution from equation  \ref{eq:zetaZ} renders the  $\chi_\nu^2$  function for a constant $\lambda_{Z^\prime}$  virtually linear,  with  $V_i$  decreasing proportional to $m_{Z^\prime}^{-2}$.
			
\item  Conversely, the effect amplifies for augmented values of $\lambda_{Z^\prime}$ (or reduced values of $m_{Z^\prime}$), until it reaches a $\chi_\nu^2$ saturation function. 
\end{enumerate}
These results follow directly from the contribution of $W_p$ to  $\zeta_{Z^\prime}(r)$, as specified in equation~\ref{eq:zetaZ}. 
		Figure~\ref{fig:LambdaBoson} illustrates that the magnitude of $W_p$, 
		given by equation~\ref{eq:nisphere}, diminishes as $\lambda_{Z^\prime}$ 
		decreases,  a consequence of the finite interaction range imposed by 
		the mediator mass, whilst approaching a constant value in the limit 
		of large $\lambda_{Z^\prime}$, where the Compton wavelength of the 
		$Z^\prime$ boson substantially exceeds the solar radius. For completeness, 
		we also display the magnitude of $W_e(r)$, although this function does 
		not enter into the present calculation.

		\medskip\noindent
		We found that  NSI models with $\chi_\nu^2\le 3.1$ are viable candidates for explaining solar neutrino data, as they perform at least as well as the standard neutrino flavour oscillation model (see Figure \ref{fig:NusundataChi2}). These models are characterized by $\eta_o$  values between $- 4.8$ and $0$ and  $\lambda_{Z^\prime}/R_\odot \ge 5  $. The best-fitting model has $\chi_\nu^2= 2.7$  for $\eta_o=-2$   and $\lambda_{Z^\prime}/R_\odot \ge 5$,
		which corresponds to a $Z^\prime$  mass of  $m_{Z^\prime} \le 5.7\times 10^{-17}\,{\rm eV}$.

		\section{Future Constraint}
		\label{sec:FutureData}

		\medskip\noindent
		It is worth considering what improvements might reasonably be expected from the forthcoming generation of solar neutrino experiments, in particular, the Jiangmen Underground Neutrino Observatory (JUNO)
			\citep{2020arXiv200611760J,2022arXiv221008437J,2023arXiv230303910A},
			and the XLZD consortium (which now unifies the XENON, LZ, and DARWIN collaborations)
			\citep{2020EPJC...80.1133D,2024NuPhB100316473B,2025EPJC...85.1192X}. The experimental landscape has evolved rather more quickly than anticipated: JUNO commenced physics data acquisition in August 2025 and has already, within a mere 59 days of operation, achieved measurements of the solar neutrino oscillation parameters $\theta_{12}$ and $\Delta m^2_{21}$ with precision surpassing all previous experiments combined by a factor of 1.6. A development that is both gratifying and rather unexpected \citep{2025arXiv251114593A}.
			
			\medskip\noindent
			Our projections had anticipated neutrino flux measurements from multiple solar reactions, and it transpires that JUNO's early performance exceeds these expectations. The collaboration's status report had foreseen a $^{8}\mathrm{B}$ flux precision below $2.5\%$ after six years of operation \citep{2024arXiv240507321S}, representing a considerable improvement upon the approximately $3.8\%$ achieved by SNO \citep[][and references therein]{2013PhRvC..88b5501A}. For the purposes of this analysis, we retain the original six-year targets: $0.15\%$ precision for $^7\mathrm{Be}$ and $3.1\%$ for $pep$ \citep{2023arXiv230303910A}, though these may well prove conservative. We also anticipate JUNO measuring CNO neutrinos, focusing on the reactions $^{15}\mathrm{O}\rightarrow {^{15}\mathrm{N}}+e^{+}+\nu_e$ and $^{13}\mathrm{N}\rightarrow {^{13}\mathrm{C}}+e^{+}+\nu_e$, whilst sensibly setting aside the rather troublesome $^{17}\mathrm{F}$ neutrinos. For $^{13}\mathrm{N}$ and $^{15}\mathrm{O}$ neutrino measurements, we assume a relative error of $20\%$ owing to their complexity. A figure that experience suggests may prove optimistic when dealing with nuclear physics at these energies. By contrast, Borexino achieved $2.7\%$ precision for $^7\mathrm{Be}$, $17\%$ for $pep$, and $30\%$ for $\mathrm{CNO}$ \citep{2018Natur.562..505B}.
			
			\medskip\noindent
			Furthermore, our analysis integrates a virtual XLZD data point alongside the extant data. All these projected data points are illustrated in Figure~\ref{fig:Pe} using yellow-grey markers. Based upon the standard neutrino oscillation model, we set these virtual data points at one quarter of the distance between current measurements and future predictions. For the XLZD data point specifically, we fix its position at one quarter of the distance between Borexino's $pp$ neutrino detection and the theoretical prediction (for $\eta_o=0$) in the analogous energy range. This approach aligns with the predictions of \citet{2020EPJC...80.1133D}, which inform our calculations. It bears mentioning that the LUX-ZEPLIN experiment achieved, in December 2025, the first statistically significant observation of solar $^{8}\mathrm{B}$ neutrinos via coherent elastic neutrino-nucleus scattering at $4.5\sigma$ significance, thereby entering what colleagues have rather evocatively termed the ``neutrino fog''. A development that bodes well for future $pp$ neutrino measurements \citep{2025arXiv251208065A}. Our analyses integrate both experimental and theoretical uncertainties, as outlined by \citet{2020MNRAS.498.1992C,2021ARNPS..71..491O}. These uncertainties are combined in quadrature, following the methodology detailed by \citet{2013PhRvD..88d5006L,2013ApJ...765...14L}.
			
		\medskip\noindent
		Figure~\ref{fig:Chi2All} demonstrates that incorporating projected JUNO 
		measurements into the analysis produces a gratifying reduction in 
		$\chi_\nu^2/d.o.f.$. For the standard neutrino flavour oscillation model 
		($\eta_o=0$), this quantity is expected to diminish from $0.45$ to $0.34$. 
		Within our preferred model class ($\lambda_{Z^\prime}/R_\odot=5$) and in 
		the region corresponding to the current best-fit model ($\eta_o=-2$), one 
		anticipates a reduction from $0.41$ to $0.27$. Whilst the improvement in 
		$\chi_\nu^2/d.o.f.$ attributable to XLZD alone appears modest, its 
		capability to measure $pp$ neutrinos at energies where the MSW effect 
		remains negligible,  that is to say, in the vacuum-dominated oscillation 
		regime below approximately 1~MeV, underscores its considerable value, 
		not least because such measurements furnish a largely model-independent 
		probe of the solar core.
		
\begin{figure}
			\centering 
			\includegraphics[scale=0.5]{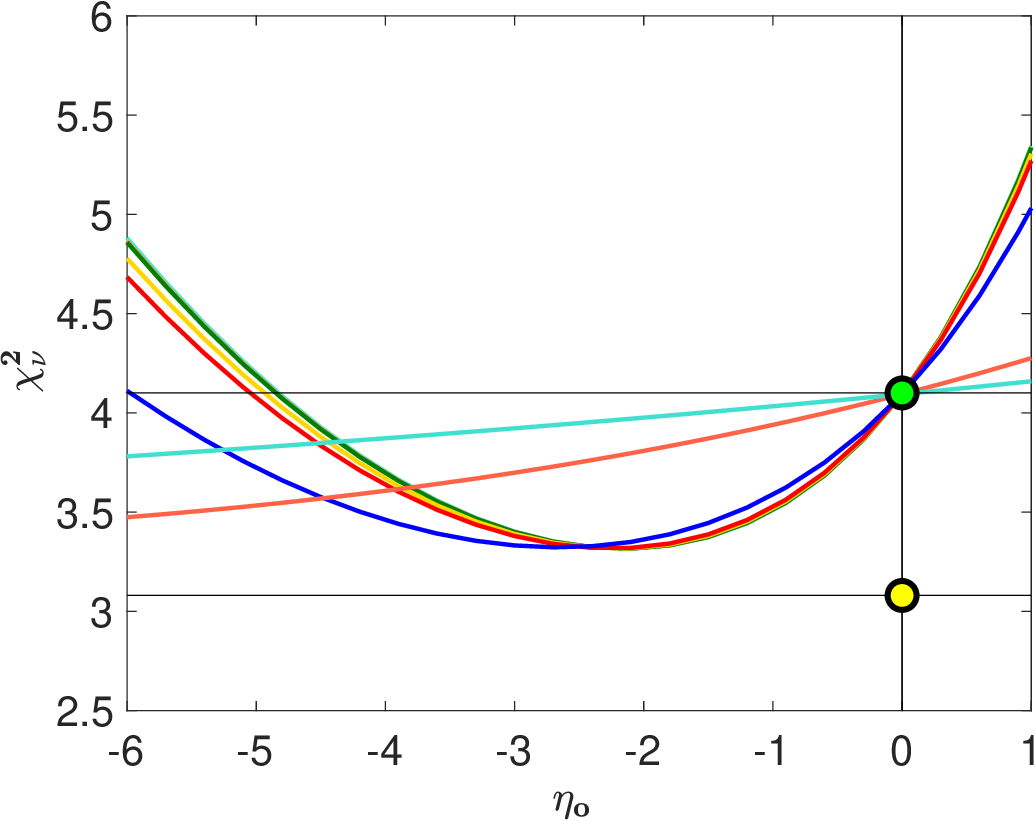}
			\caption{The $\chi^2_\nu$ test values as a function of the coupling constant $\eta_o$ for various non-standard interaction (NSI) neutrino models. This analysis incorporates all experimental data together with projected JUNO measurements (see Figure~\ref{fig:Pe}). The colour scheme follows that employed in Figure~\ref{fig:NusundataChi2}. The green circle denotes the $\chi^2_\nu$ value for the standard neutrino flavour model when evaluated against both present and anticipated JUNO data. Further particulars may be found in the main text.}
			\label{fig:Chi2Juno}
		\end{figure}

\begin{figure} 
			\includegraphics[scale=0.5]{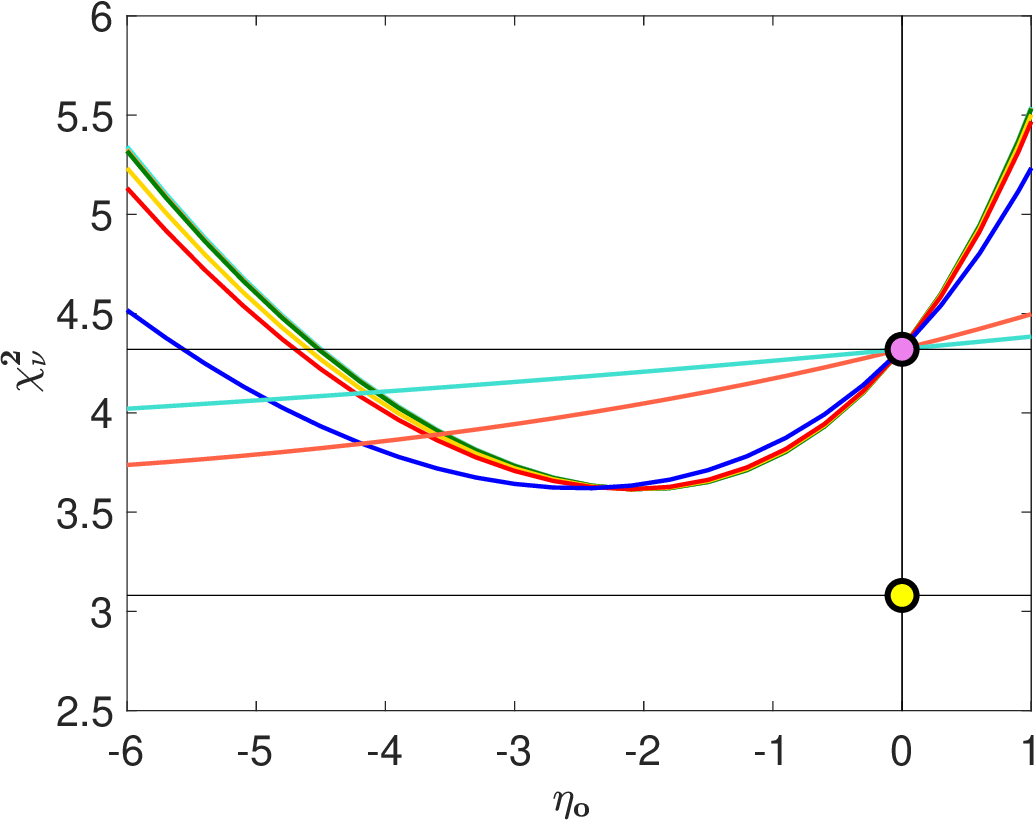}
			\caption{The $\chi^2_\nu$ test values versus the coupling constant $\eta_o$ across various non-standard interaction (NSI) neutrino models. This analysis mirrors that of Figure~\ref{fig:Chi2Juno}, but now incorporates projected data from both JUNO and XLZD (see Figure~\ref{fig:Pe}). The colour scheme follows Figure~\ref{fig:NusundataChi2}. The violet circle indicates the $\chi_\nu^2$ for the standard neutrino flavour model, when combining extant data with projected JUNO and XLZD data. Further details are provided in the main text.}
			\label{fig:Chi2DarwinJuno}
		\end{figure}

		\medskip\noindent
		In Figures~\ref{fig:Chi2Juno} and \ref{fig:Chi2DarwinJuno}, we present 
		analogous plots to Figure~\ref{fig:NusundataChi2}. These figures illustrate 
		the potential model constraints based upon six years of JUNO data, as well 
		as the combined dataset from both JUNO and XLZD experiments. For scenarios 
		where $\lambda_{Z^\prime}/R_\odot \ge 5$, the optimal model range is 
		expected to narrow from $-6.0 \le \eta_o\le 0$ (based upon current data) 
		to $-4.5 \le \eta_o\le 0$ when incorporating the full JUNO and XLZD dataset.

		\section{Summary and Conclusion} 
		\label{sec-Con}

\noindent
In this letter, we have presented what one hopes is a nuanced adaptation of the traditional particle physics model, crafted to address observed phenomena in solar neutrino physics. Central to this framework is the gauged $U(1)_{L_\mu - L_\tau}$ symmetry, a naturally anomaly-free extension of the Standard Model that couples to the difference of muon and tau lepton numbers. 
We theorise that the $Z^\prime$ boson interacts with the solar plasma through kinetic mixing with the photon, thereby generating an effective matter potential for neutrinos. This addition complements (rather than supplants) the conventional three-neutrino flavour oscillation framework. Our preliminary results suggest that this model matches the current benchmark set by the standard neutrino flavour oscillation model in fitting solar neutrino data, and may indeed exceed it. Though one must, of course, remain appropriately cautious about such claims.
	
\medskip\noindent
The $L_\mu - L_\tau$ gauge symmetry was historically motivated by the anomalous magnetic dipole moment of the muon and related flavour anomalies \citep{2011JPhG...38h5005H}. It is important to acknowledge, in the interest of scientific candour, that the landscape regarding these anomalies has been fundamentally transformed. The B-meson anomalies in $R_K$ and $R_{K^*}$, which had excited considerable theoretical speculation, were resolved following improved background modelling by the LHCb collaboration in December 2022 \citep{2023EPJC...83..648A}. More significantly, the Fermilab Muon g-2 experiment released its final measurement in June 2025 \citep{2025PhRvL.135j1802A}, achieving an unprecedented precision of 127 parts per billion. Concurrently, the Muon g-2 Theory Initiative published an updated Standard Model prediction \citep{2025PhR..1143....1A} incorporating mature lattice QCD calculations for the hadronic vacuum polarisation contribution. The resulting comparison reveals no statistically significant tension between theory and experiment at current precision levels. These developments, whilst perhaps disappointing for enthusiasts of new physics via the g-2 route, do not diminish the independent validity of our model when constrained by solar neutrino data. Indeed, they underscore the importance of diverse experimental probes such as the solar neutrino measurements we employ here, for testing extensions of the Standard Model. The $L_\mu - L_\tau$ model remains a highly attractive candidate for future study due to its ability to address terrestrial long-baseline force searches \citep{2023JHEP...08..101S,2024JHEP...09..055A} and non-thermal dark matter phenomenology \citep{2025PhRvD.111c5008Q}.

\medskip\noindent
Employing the most recent solar neutrino data in conjunction with the current solar standard model, we have placed constraints on this Non-Standard Interaction model. Our results indicate that the best agreement with solar neutrino data arises when the intermediate boson is ultra-light. In particular, the mass of the $Z^\prime$ boson should be less than $5.7\times 10^{-17}\,{\rm eV}$ (or equivalently, $\lambda_{Z^\prime}/R_\odot \ge 5$) for optimal alignment with current solar neutrino measurements.

\medskip\noindent
Additionally, models with $-6.0 \le \eta_o \le 0$ achieve a fit to the current 
solar neutrino data that is comparable to, and in some cases marginally better 
than the standard flavour oscillation scenario. It is worth remarking that the 
best-fit model corresponds to a coupling parameter $\eta_{o} = g_{Z^\prime}^2 
\varepsilon / (2\sqrt{2}\, G_F\, m_{Z^\prime}^2) \approx -2$, where $g_{Z^\prime}$ denotes the 
gauge coupling constant, $\varepsilon$ the kinetic mixing parameter, and $G_F$ 
the Fermi constant. This model yields $\chi_\nu^2 = 2.7$, which stands in contrast 
to the $\chi_\nu^2 = 3.1$ obtained for the standard case ($\eta_o = 0$). The value 
$\eta_{o} = -2$ corresponds to an effective four-fermion coupling $g_{Z^\prime}^2 \varepsilon / m_{Z^\prime}^2 = 2\sqrt{2}\, G_F |\eta_o| \approx 6.6 \times 10^{-23}\;\mathrm{eV}^{-2}$, that is, approximately $5.7$ times the Fermi constant, which is precisely why the induced potential is able to compete with the standard MSW term. Assuming $\varepsilon \approx 1$ and adopting a reference mediator mass $m_{Z^\prime} = 1\,\mathrm{eV}$, this corresponds to a gauge coupling $g_{Z^\prime} \approx 8\times 10^{-12}$, comfortably within the range anticipated for ultra-light mediators (see below). It bears noting that 
whilst the individual values of the gauge coupling $g_{Z^\prime}$ and kinetic 
mixing parameter $\varepsilon$ are not separately constrained by this analysis, 
existing bounds from neutrino scattering experiments and cosmological considerations 
suggest that both parameters may span several orders of magnitude. For instance, 
cosmological constraints permit $g_{Z^\prime} \simeq (3\text{--}8) \times 10^{-4}$ 
for MeV-scale $Z^\prime$ bosons, whilst for ultra-light mediators ($m_{Z^\prime} 
\sim 1\;\mathrm{eV}$ to $\mathrm{MeV}$) the coupling may lie in the range 
$g_{Z^\prime} \sim 10^{-13}$ to $10^{-9}$ \citep{2019JHEP...03..071E}. Similarly, 
the kinetic mixing parameter $\varepsilon$ is scale-dependent and may effectively 
vanish at low momentum transfer \citep{2022PhRvD.105a6014H}. The product 
$g_{Z^\prime}^2 \varepsilon$ nonetheless remains the physically relevant 
combination for solar neutrino phenomenology.

\medskip\noindent
Furthermore, our findings suggest that the forthcoming generation of solar neutrino detectors, JUNO and XLZD in particular, can significantly refine $\chi_\nu^2/d.o.f.$. Indeed, we anticipate that the standard case's value of $0.45$ may be reduced to $0.34$ when utilising combined data from these collaborations. This advancement paves the way for either validating the standard scenario, or confirming this (or similar) Non-Standard Interaction models. Given that JUNO has already commenced operations and delivered results of remarkable precision, one is cautiously optimistic that these projections may be realised sooner than originally anticipated.

\medskip\noindent 
The implication that the intermediary boson is an ultra-light particle, with mass constrained to be less than $5.7\times 10^{-17}\,{\rm eV}$, suggests that it may, in principle, address several outstanding challenges in astrophysics and cosmology. We highlight a few scenarios potentially linked to such particles:
\begin{itemize} 
\item[--]
If stable, this boson emerges as a viable dark matter candidate. Ultra-light dark matter particles offer notable potential in refining galaxy simulations within the CDM paradigm. Such particles serve as a bridge to resolve numerous discrepancies between numerical predictions and cosmological observations, including the core versus cusp issue, the missing satellites dilemma, the too-big-to-fail problem, galaxy rotation curve anomalies, and complications in high-redshift galaxy formation \citep[e.g.,][]{2021A&ARv..29....7F,2021ARA&A..59..247H}. Recent constraints from dwarf galaxy kinematics suggest that ultra-light bosonic dark matter must have masses exceeding approximately $2.2\times 10^{-21}\,{\rm eV}$ \citep{2025PhRvL.134o1001Z}; our candidate, being considerably lighter still, occupies a distinct region of parameter space that would require different phenomenological signatures. Importantly, recent theoretical developments have demonstrated that the $L_\mu - L_\tau$ model can accommodate non-thermal dark matter production via the decay of heavy right-handed neutrinos \citep{2025PhRvD.111c5008Q}, potentially explaining both the dark matter relic density and the baryon asymmetry of the Universe through leptogenesis.

\item[--] 
It is worth remarking that the kinetic mixing inherent to this model gives rise to 
long-range forces that are, in principle, constrained by tests of the weak 
equivalence principle. Through kinetic mixing with the photon, the $Z^\prime$ boson 
acquires an effective coupling to the electromagnetic current proportional to 
$\varepsilon e$, which can mediate composition-dependent interactions between 
macroscopic bodies. The MICROSCOPE satellite experiment has placed stringent 
bounds on new long-range forces coupled to linear combinations of baryon and 
lepton numbers, with constraints at the level of $|\varepsilon| \lesssim 10^{-24}$ 
for forces coupled to total lepton number $L$ or $B - L$ 
\citep{2019PhRvD..99e5043F}. For the $L_\mu - L_\tau$ model under consideration 
here, however, neither electrons nor quarks carry any direct charge under the 
gauge symmetry, and consequently fifth-force signatures arise only through the 
kinetic mixing portal, a qualitatively different and suppressed mechanism 
compared to models such as $L_e - L_\mu$ or $L_e - L_\tau$, where electrons 
couple directly to the new gauge boson and planetary perihelion precession  
measurements yield constraints of order $g \lesssim 10^{-25}$ 
\citep{2021EPJC...81..286P}. Importantly, the \emph{physical} kinetic mixing 
relevant for low-energy processes is not simply the bare Lagrangian parameter 
$\varepsilon$, but rather receives radiative corrections from $\mu$ and $\tau$ 
loops that render it momentum-dependent. In the low-momentum regime pertinent to 
fifth-force experiments, this effective mixing approaches an infrared boundary 
value that constitutes a free parameter of the model 
\citep{2022PhRvD.105a6014H}. If this boundary value lies in proximity of zero, 
constraints from solar neutrino experiments, coherent elastic neutrino, nucleus 
scattering, and stellar cooling can be substantially relaxed, potentially by 
orders of magnitude. Nevertheless, any viable parameter space must remain 
consistent with these complementary probes. This phenomenological situation is 
distinct from baryon-coupled models, where the electron sector plays no 
mediating role whatsoever.

\item[--] The introduction of a new ultra-light boson necessitates re-evaluating pivotal processes in stellar evolution. 
One anticipates that nuclear reactions, the equation of state, and opacities in the Sun and other stars are amongst these processes. This perspective could provide critical insights into the solar abundance problem \citep[e.g.,][]{2005ApJ...621L..85B,2008PhR...457..217B}, a persistent inconsistency amongst solar models, observed neutrino fluxes, and helioseismic observations that has resisted resolution despite considerable effort over the past two decades. Such alterations hold promise for predicting a seismic structure more consistent with observational data.

\item[--] Complementary tests of this model arise from terrestrial long-baseline neutrino experiments. Studies by \citet{2023JHEP...08..101S} and \citet{2024JHEP...09..055A} have demonstrated that a combined analysis of DUNE and T2HK data is required to claim a statistical discovery of flavour-dependent long-range forces. Furthermore, high-energy astrophysical neutrinos offer a lever arm to probe coupling constants as small as $10^{-30}$, far below solar sensitivity.
\end{itemize}

\noindent 
Finally, the forthcoming generation of ultra-light dark matter detectors, as detailed 
in reviews such as \citet{2019SciA....5.4539G}, \citet{2021EPJC...81.1015A}, and 
\citet{2021A&ARv..29....7F} is poised to probe this class of particle with increasing 
rigour. The bosons discussed in this work, characterised by masses below 
$5.7\times 10^{-17}\,{\rm eV}$ and an effective four-fermion coupling $g_{Z^\prime}^2\varepsilon/m_{Z^\prime}^2 \approx 6.6\times 10^{-23}\;{\rm eV}^{-2}$ (about $5.7\,G_F$), represent a compelling target for experimental 
investigation. It bears noting, however, that such extraordinarily light mediators 
lie at the frontier of current detection capabilities, with gravitational-wave 
interferometers, pulsar timing arrays, and precision atomic spectroscopy amongst the 
most promising avenues for exploration \citep[see, e.g.,][]{2019PhRvL.123c1304G, 
	2021Natur.600..424V}.
The detection of ultra-light bosons and further theoretical refinements could furnish 
important insights into the interplay between dark matter and leptonic interactions, 
potentially reshaping our understanding of the Universe's underlying structure. One 
remains, as ever, cautiously hopeful.

\section*{Acknowledgments}

\medskip\noindent
I.L. would like to express his gratitude to the Funda\c c\~ao para a Ci\^encia e Tecnologia (FCT), Portugal, for providing financial support to the Center for Astrophysics and Gravitation (CENTRA/IST/ULisboa) through Grant Project No. UID/PRR/00099/2025  (https://doi.org/10.54499/UID/PRR/00099/2025) and Grant Project No.  UID/00099/2025  (https://doi.org/10.54499/UID/00099/2025).

		\renewcommand{\bibfont}{\footnotesize}
		\bibliography{main}

@ARTICLE{2019PhRvD..99e5043F,
	author = {{Fayet}, Pierre},
	title = "{M I C R O S C O P E limits on the strength of a new force with comparisons to gravity and electromagnetism}",
	journal = {\prd},
	year = 2019,
	month = mar,
	volume = {99},
	number = {5},
	eid = {055043},
	pages = {055043},
	url = {https://doi.org/10.1103/PhysRevD.99.055043},
	doi = {10.1103/PhysRevD.99.055043},
	archivePrefix = {arXiv},
	eprint = {1809.04991},
	primaryClass = {hep-ph},
	adsurl = {https://ui.adsabs.harvard.edu/abs/2019PhRvD..99e5043F}
}

@ARTICLE{2021Natur.600..424V,
	author = {{Vermeulen}, Sander M. and {Relton}, Philip and {Grote}, Hartmut and {Raymond}, Vivien and {Affeldt}, Christoph and {Bergamin}, Fabio and {Bisht}, Aparna and {Brinkmann}, Marc and {Danzmann}, Karsten and {Doravari}, Suresh and {Kringel}, Volker and {Lough}, James and {L{\"u}ck}, Harald and {Mehmet}, Moritz and {Mukund}, Nikhil and {Nadji}, S{\'e}verin and {Schreiber}, Emil and {Sorazu}, Borja and {Strain}, Kenneth A. and {Vahlbruch}, Henning and {Weinert}, Michael and {Willke}, Benno and {Wittel}, Holger},
	title = "{Direct limits for scalar field dark matter from a gravitational-wave detector}",
	journal = {\nat},
	year = 2021,
	month = dec,
	volume = {600},
	number = {7889},
	pages = {424-428},
	url = {https://doi.org/10.1038/s41586-021-04031-y},
	doi = {10.1038/s41586-021-04031-y},
	archivePrefix = {arXiv},
	eprint = {2103.03783},
	primaryClass = {gr-qc},
	adsurl = {https://ui.adsabs.harvard.edu/abs/2021Natur.600..424V}
}

@ARTICLE{2019PhRvL.123c1304G,
	author = {{Geraci}, Andrew A. and {Bradley}, Colin and {Gao}, Dongfeng and {Weinstein}, Jonathan and {Derevianko}, Andrei},
	title = "{Searching for Ultralight Dark Matter with Optical Cavities}",
	journal = {\prl},
	year = 2019,
	month = jul,
	volume = {123},
	number = {3},
	eid = {031304},
	pages = {031304},
	url = {https://doi.org/10.1103/PhysRevLett.123.031304},
	doi = {10.1103/PhysRevLett.123.031304},
	archivePrefix = {arXiv},
	eprint = {1808.00540},
	primaryClass = {astro-ph.IM},
	adsurl = {https://ui.adsabs.harvard.edu/abs/2019PhRvL.123c1304G}
}

@ARTICLE{2022PhRvD.105a6014H,
	author = {{Hapitas}, Timothy and {Tuckler}, Douglas and {Zhang}, Yue},
	title = "{General kinetic mixing in gauged U$_{(1 ) L<SUB>{\ensuremath{\mu}}}$-L$_{{\ensuremath{\tau}}}$</SUB> model for muon g -2 and dark matter}",
	journal = {\prd},
	year = 2022,
	month = jan,
	volume = {105},
	number = {1},
	eid = {016014},
	pages = {016014},
	url = {https://doi.org/10.1103/PhysRevD.105.016014},
	doi = {10.1103/PhysRevD.105.016014},
	archivePrefix = {arXiv},
	eprint = {2108.12440},
	primaryClass = {hep-ph},
	adsurl = {https://ui.adsabs.harvard.edu/abs/2022PhRvD.105a6014H}
}

@ARTICLE{2019JHEP...03..071E,
	author = {{Escudero}, Miguel and {Hooper}, Dan and {Krnjaic}, Gordan and {Pierre}, Mathias},
	title = "{Cosmology With a Very Light $L_\mu - L_\tau$ Gauge Boson}",
	journal = {arXiv e-prints},
	year = 2019,
	month = jan,
	eid = {arXiv:1901.02010},
	pages = {arXiv:1901.02010},
	url = {https://doi.org/10.48550/arXiv.1901.02010},
	doi = {10.48550/arXiv.1901.02010},
	archivePrefix = {arXiv},
	eprint = {1901.02010},
	primaryClass = {hep-ph},
	adsurl = {https://ui.adsabs.harvard.edu/abs/2019arXiv190102010E}
}

@ARTICLE{2011JPhG...38h5005H,
	author = {{Heeck}, Julian and {Rodejohann}, Werner},
	title = "{Gauged L$_{{\ensuremath{\mu}}}$-L$_{{\ensuremath{\tau}}}$ and different muon neutrino and anti-neutrino oscillations: MINOS and beyond}",
	journal = {Journal of Physics G Nuclear Physics},
	year = 2011,
	month = aug,
	volume = {38},
	number = {8},
	eid = {085005},
	pages = {085005},
	url = {https://doi.org/10.1088/0954-3899/38/8/085005},
	doi = {10.1088/0954-3899/38/8/085005},
	archivePrefix = {arXiv},
	eprint = {1007.2655},
	primaryClass = {hep-ph},
	adsurl = {https://ui.adsabs.harvard.edu/abs/2011JPhG...38h5005H}
}

@ARTICLE{2025PhRvD.111c5008Q,
	author = {{Qi}, XinXin and {Sun}, Hao},
	title = "{Nonthermal dark matter production and leptogenesis in a L{\ensuremath{\mu}}-L{\ensuremath{\tau}} model}",
	journal = {\prd},
	year = 2025,
	month = feb,
	volume = {111},
	number = {3},
	eid = {035008},
	pages = {035008},
	url = {https://doi.org/10.1103/PhysRevD.111.035008},
	doi = {10.1103/PhysRevD.111.035008},
	archivePrefix = {arXiv},
	eprint = {2204.01086},
	primaryClass = {hep-ph},
	adsurl = {https://ui.adsabs.harvard.edu/abs/2025PhRvD.111c5008Q}
}

@ARTICLE{2023JHEP...08..101S,
	author = {{Singh}, Masoom and {Bustamante}, Mauricio and {Agarwalla}, Sanjib Kumar},
	title = "{Flavor-dependent long-range neutrino interactions in DUNE \& T2HK: alone they constrain, together they discover}",
	journal = {Journal of High Energy Physics},
	year = 2023,
	month = aug,
	volume = {2023},
	number = {8},
	eid = {101},
	pages = {101},
	url = {https://doi.org/10.1007/JHEP08(2023)101},
	doi = {10.1007/JHEP08(2023)101},
	archivePrefix = {arXiv},
	eprint = {2305.05184},
	primaryClass = {hep-ph},
	adsurl = {https://ui.adsabs.harvard.edu/abs/2023JHEP...08..101S}
}

@ARTICLE{2024JHEP...09..055A,
	author = {{Agarwalla}, Sanjib Kumar and {Bustamante}, Mauricio and {Singh}, Masoom and {Swain}, Pragyanprasu},
	title = "{A plethora of long-range neutrino interactions probed by DUNE and T2HK}",
	journal = {Journal of High Energy Physics},
	year = 2024,
	month = sep,
	volume = {2024},
	number = {9},
	eid = {55},
	pages = {55},
	url = {https://doi.org/10.1007/JHEP09(2024)055},
	doi = {10.1007/JHEP09(2024)055},
	archivePrefix = {arXiv},
	eprint = {2404.02775},
	primaryClass = {hep-ph},
	adsurl = {https://ui.adsabs.harvard.edu/abs/2024JHEP...09..055A}
}

@ARTICLE{2021EPJC...81..286P,
	author = {{Poddar}, Tanmay Kumar and {Mohanty}, Subhendra and {Jana}, Soumya},
	title = "{Constraints on long range force from perihelion precession of planets in a gauged L$_{e}$-L$_{{\ensuremath{\mu}} ,{\ensuremath{\tau}}}$ scenario}",
	journal = {European Physical Journal C},
	year = 2021,
	month = apr,
	volume = {81},
	number = {4},
	eid = {286},
	pages = {286},
	url = {https://doi.org/10.1140/epjc/s10052-021-09078-9},
	doi = {10.1140/epjc/s10052-021-09078-9},
	archivePrefix = {arXiv},
	eprint = {2002.02935},
	primaryClass = {hep-ph},
	adsurl = {https://ui.adsabs.harvard.edu/abs/2021EPJC...81..286P}
}

@ARTICLE{2023JHEP...08..032C,
	author = {{Coloma}, Pilar and {Gonzalez-Garcia}, M.~C. and {Maltoni}, Michele and {Pinheiro}, Jo{\~a}o Paulo and {Urrea}, Salvador},
	title = "{Global constraints on non-standard neutrino interactions with quarks and electrons}",
	journal = {Journal of High Energy Physics},
	year = 2023,
	month = aug,
	volume = {2023},
	number = {8},
	eid = {32},
	pages = {32},
	url = {https://doi.org/10.1007/JHEP08(2023)032},
	doi = {10.1007/JHEP08(2023)032},
	archivePrefix = {arXiv},
	eprint = {2305.07698},
	primaryClass = {hep-ph},
	adsurl = {https://ui.adsabs.harvard.edu/abs/2023JHEP...08..032C}
}

@ARTICLE{2019arXiv190700991B,
	author = {{Bhupal Dev}, P.~S. and {Babu}, K.~S. and {Denton}, Peter B. and {Machado}, Pedro A.~N. and {Arg{\"u}elles}, Carlos A. and {Barrow}, Joshua L. and {Sachi Chatterjee}, Sabya and {Chen}, Mu-Chun and {de Gouv{\^e}a}, Andr{\'e} and {Dutta}, Bhaskar and {Gon{\c{c}}alves}, Dorival and {Han}, Tao and {Hostert}, Matheus and {Jana}, Sudip and {Kelly}, Kevin J. and {Weishi Li}, Shirley and {Martinez-Soler}, Ivan and {Mehta}, Poonam and {Mocioiu}, Irina and {Perez-Gonzalez}, Yuber F. and {Salvado}, Jordi and {Shoemaker}, Ian M. and {Tammaro}, Michele and {Thapa}, Anil and {Turner}, Jessica and {Xu}, Xun-Jie},
	title = "{Neutrino Non-Standard Interactions: A Status Report}",
	journal = {arXiv e-prints},
	year = 2019,
	month = jul,
	eid = {arXiv:1907.00991},
	pages = {arXiv:1907.00991},
	url = {https://doi.org/10.48550/arXiv.1907.00991},
	doi = {10.48550/arXiv.1907.00991},
	archivePrefix = {arXiv},
	eprint = {1907.00991},
	primaryClass = {hep-ph},
	adsurl = {https://ui.adsabs.harvard.edu/abs/2019arXiv190700991B}
}

@ARTICLE{2025PhR..1143....1A,
	author = {{Aliberti}, R. and {Aoyama}, T. and {Balzani}, E. and {Bashir}, A. and {Benton}, G. and {Bijnens}, J. and {Biloshytskyi}, V. and {Blum}, T. and {Boito}, D. and {Bruno}, M. and {Budassi}, E. and {Burri}, S. and {Cappiello}, L. and {Carloni Calame}, C.~M. and {C{\`e}}, M. and {Cirigliano}, V. and {Clarke}, D.~A. and {Colangelo}, G. and {Cotrozzi}, L. and {Cottini}, M. and {Danilkin}, I. and {Davier}, M. and {Della Morte}, M. and {Denig}, A. and {DeTar}, C. and {Druzhinin}, V. and {Eichmann}, G. and {El-Khadra}, A.~X. and {Estrada}, E. and {Feng}, X. and {Fischer}, C.~S. and {Frezzotti}, R. and {Gagliardi}, G. and {G{\'e}rardin}, A. and {Ghilardi}, M. and {Giusti}, D. and {Golterman}, M. and {Gonz{\`a}lez-Sol{\'\i}s}, S. and {Gottlieb}, S. and {Gruber}, R. and {Guevara}, A. and {G{\"u}lpers}, V. and {Gurgone}, A. and {Hagelstein}, F. and {Hayakawa}, M. and {Hermansson-Truedsson}, N. and {Hoecker}, A. and {Hoferichter}, M. and {Hoid}, B.-L. and {Holz}, S. and {Hudspith}, R.~J. and {Ignatov}, F. and {Jin}, L. and {Kalntis}, N. and {Kanwar}, G. and {Keshavarzi}, A. and {Komijani}, J. and {Koponen}, J. and {Kuberski}, S. and {Kubis}, B. and {Kupich}, A. and {Kup{\'s}{\'c}}, A. and {Lahert}, S. and {Laporta}, S. and {Lehner}, C. and {Lellmann}, M. and {Lellouch}, L. and {Leplumey}, T. and {Leutgeb}, J. and {Lin}, T. and {Liu}, Q. and {Logashenko}, I. and {London}, C.~Y. and {L{\'o}pez Castro}, G. and {L{\"u}dtke}, J. and {Lusiani}, A. and {Lutz}, A. and {Mager}, J. and {Malaescu}, B. and {Maltman}, K. and {Marinkovi{\'c}}, M.~K. and {M{\'a}rquez}, J. and {Masjuan}, P. and {Meyer}, H.~B. and {Mibe}, T. and {Miller}, N. and {Miramontes}, A. and {Miranda}, A. and {Montagna}, G. and {M{\"u}ller}, S.~E. and {Neil}, E.~T. and {Nesterenko}, A.~V. and {Nicrosini}, O. and {Nio}, M. and {Nomura}, D. and {Paltrinieri}, J. and {Parato}, L. and {Parrino}, J. and {Pascalutsa}, V. and {Passera}, M. and {Peris}, S. and {Petit Ros{\`a}s}, P. and {Piccinini}, F. and {Pilato}, R.~N. and {Polat}, L. and {Portelli}, A. and {Portillo-S{\'a}nchez}, D. and {Procura}, M. and {Punzi}, L. and {Raya}, K. and {Rebhan}, A. and {Redmer}, C.~F. and {Roberts}, B.~L. and {Rodr{\'\i}guez-S{\'a}nchez}, A. and {Roig}, P. and {Ruiz de Elvira}, J. and {S{\'a}nchez-Puertas}, P. and {Signer}, A. and {Sitison}, J.~W. and {Stamen}, D. and {St{\"o}ckinger}, D. and {St{\"o}ckinger-Kim}, H. and {Stoffer}, P. and {Sue}, Y. and {Tavella}, P. and {Teubner}, T. and {Toelstede}, J.-N. and {Toledo}, G. and {Torres Bobadilla}, W.~J. and {Tsang}, J.~T. and {Ucci}, F.~P. and {Ulrich}, Y. and {Van de Water}, R.~S. and {Venanzoni}, G. and {Volkov}, S. and {von Hippel}, G. and {Wang}, G. and {Wenger}, U. and {Wittig}, H. and {Wright}, A. and {Zaid}, E. and {Zanke}, M. and {Zhang}, Z. and {Zillinger}, M. and {Alexandrou}, C. and {Altherr}, A. and {Anderson}, M. and {Aubin}, C. and {Bacchio}, S. and {Beltrame}, P. and {Beltran}, A. and {Boyle}, P. and {Campos Plasencia}, I. and {Caprini}, I. and {Chakraborty}, B. and {Chanturia}, G. and {Crivellin}, A. and {Czarnecki}, A. and {Dai}, L.-Y. and {Dave}, T. and {Del Debbio}, L. and {Demory}, K. and {Djukanovic}, D. and {Draper}, T. and {Driutti}, A. and {Endo}, M. and {Erben}, F. and {Ferraby}, K. and {Finkenrath}, J. and {Flower}, L. and {Francis}, A. and {G{\'a}miz}, E. and {Gogniat}, J. and {Grebe}, A.~V. and {G{\"u}ndogdu}, S. and {Hansen}, M.~T. and {Hashimoto}, S. and {Hayashii}, H. and {Hertzog}, D.~W. and {Heuser}, L.~A. and {Hostetler}, L. and {Hou}, X.~T. and {Huang}, G.~S. and {Iijima}, T. and {Inami}, K. and {J{\"u}ttner}, A. and {Kitano}, R. and {Knecht}, M. and {Kollatzsch}, S. and {Kronfeld}, A.~S. and {Lenz}, T. and {Levati}, G. and {Li}, Q.~M. and {Liao}, Y.~P. and {Libby}, J. and {Liu}, K.~F. and {Lubicz}, V. and {Lynch}, M.~T. and {Lytle}, A.~T. and {Ma}, J.~L.},
	title = "{The anomalous magnetic moment of the muon in the Standard Model: an update}",
	journal = {\physrep},
	year = 2025,
	month = nov,
	volume = {1143},
	pages = {1-158},
	url = {https://doi.org/10.1016/j.physrep.2025.08.002},
	doi = {10.1016/j.physrep.2025.08.002},
	archivePrefix = {arXiv},
	eprint = {2505.21476},
	primaryClass = {hep-ph},
	adsurl = {https://ui.adsabs.harvard.edu/abs/2025PhR..1143....1A}
}

@ARTICLE{2025PhRvL.135j1802A,
	author = {{Aguillard}, D.~P. and {Albahri}, T. and {Allspach}, D. and {Annala}, J. and {Badgley}, K. and {Bae{\ss}ler}, S. and {Bailey}, I. and {Bailey}, L. and {Barlas-Yucel}, E. and {Barrett}, T. and {Barzi}, E. and {Bedeschi}, F. and {Berz}, M. and {Bhattacharya}, M. and {Binney}, H.~P. and {Bloom}, P. and {Bono}, J. and {Bottalico}, E. and {Bowcock}, T. and {Braun}, S. and {Bressler}, M. and {Cantatore}, G. and {Carey}, R.~M. and {Casey}, B.~C.~K. and {Cauz}, D. and {Chakraborty}, R. and {Chapelain}, A. and {Chappa}, S. and {Charity}, S. and {Chen}, C. and {Cheng}, M. and {Chislett}, R. and {Chu}, Z. and {Chupp}, T.~E. and {Claessens}, C. and {Confortini}, F. and {Convery}, M.~E. and {Corrodi}, S. and {Cotrozzi}, L. and {Crnkovic}, J.~D. and {Dabagov}, S. and {Debevec}, P.~T. and {Di Falco}, S. and {Di Sciascio}, G. and {Donati}, S. and {Drendel}, B. and {Driutti}, A. and {Eads}, M. and {Edmonds}, A. and {Esquivel}, J. and {Farooq}, M. and {Fatemi}, R. and {Ferraby}, K. and {Ferrari}, C. and {Fertl}, M. and {Fienberg}, A.~T. and {Fioretti}, A. and {Flay}, D. and {Foster}, S.~B. and {Friedsam}, H. and {Froemming}, N.~S. and {Gabbanini}, C. and {Gaines}, I. and {Ganguly}, S. and {George}, J. and {Gibbons}, L.~K. and {Gioiosa}, A. and {Giovanetti}, K.~L. and {Girotti}, P. and {Gohn}, W. and {Goodenough}, L. and {Gorringe}, T. and {Grange}, J. and {Grant}, S. and {Gray}, F. and {Haciomeroglu}, S. and {Halewood-Leagas}, T. and {Hampai}, D. and {Han}, F. and {Hempstead}, J. and {Hertzog}, D.~W. and {Hesketh}, G. and {Hess}, E. and {Hibbert}, A. and {Hodge}, Z. and {Hoh}, S.~Y. and {Hong}, K.~W. and {Hong}, R. and {Hu}, T. and {Hu}, Y. and {Iacovacci}, M. and {Incagli}, M. and {Israel}, S. and {Kammel}, P. and {Kargiantoulakis}, M. and {Karuza}, M. and {Kaspar}, J. and {Kawall}, D. and {Kelton}, L. and {Keshavarzi}, A. and {Kessler}, D.~S. and {Khaw}, K.~S. and {Khechadoorian}, Z. and {Kiburg}, B. and {Kiburg}, M. and {Kim}, O. and {Kinnaird}, N. and {Kraegeloh}, E. and {Labe}, K.~R. and {LaBounty}, J. and {Lancaster}, M. and {Lee}, S. and {Li}, B. and {Li}, D. and {Li}, L. and {Logashenko}, I. and {Lorente Campos}, A. and {Lu}, Z. and {Luc{\`a}}, A. and {Lukicov}, G. and {Lusiani}, A. and {Lyon}, A.~L. and {MacCoy}, B. and {Madrak}, R. and {Makino}, K. and {Mastroianni}, S. and {McCarthy}, R. and {Miller}, J.~P. and {Miozzi}, S. and {Mitra}, B. and {Morgan}, J.~P. and {Morse}, W.~M. and {Mott}, J. and {Nath}, A. and {Ng}, J.~K. and {Nguyen}, H. and {Oksuzian}, Y. and {Omarov}, Z. and {Osar}, W. and {Osofsky}, R. and {Park}, S. and {Pauletta}, G. and {Peck}, J. and {Piacentino}, G.~M. and {Pilato}, R.~N. and {Pitts}, K.~T. and {Plaster}, B. and {Po{\v{c}}ani{\'c}}, D. and {Pohlman}, N. and {Polly}, C.~C. and {Price}, J. and {Quinn}, B. and {Qureshi}, M.~U.~H. and {Rakness}, G. and {Ramachandran}, S. and {Ramberg}, E. and {Reimann}, R. and {Roberts}, B.~L. and {Rubin}, D.~L. and {Sakurai}, M. and {Santi}, L. and {Schlesier}, C. and {Schreckenberger}, A. and {Semertzidis}, Y.~K. and {Soha}, A.~K. and {Sorbara}, M. and {Stapleton}, J. and {Still}, D. and {St{\"o}ckinger}, D. and {Stoughton}, C. and {Stratakis}, D. and {Swanson}, H.~E. and {Sweetmore}, G. and {Sweigart}, D.~A. and {Syphers}, M.~J. and {Takeuchi}, Y. and {Tarazona}, D.~A. and {Teubner}, T. and {Tewsley-Booth}, A.~E. and {Tishchenko}, V. and {Tran}, N.~H. and {Turner}, W. and {Valetov}, E. and {Vasilkova}, D. and {Venanzoni}, G. and {Walton}, T. and {Weisskopf}, A. and {Welty-Rieger}, L. and {Winter}, P. and {Wu}, Y. and {Yu}, B. and {Yucel}, M. and {Zaid}, E. and {Zeng}, Y. and {Zhang}, C. and {(Muon}},
	title = "{Measurement of the Positive Muon Anomalous Magnetic Moment to 127 ppb}",
	journal = {\prl},
	year = 2025,
	month = sep,
	volume = {135},
	number = {10},
	eid = {101802},
	pages = {101802},
	url = {https://doi.org/10.1103/7clf-sm2v},
	doi = {10.1103/7clf-sm2v},
	archivePrefix = {arXiv},
	eprint = {2506.03069},
	primaryClass = {hep-ex},
	adsurl = {https://ui.adsabs.harvard.edu/abs/2025PhRvL.135j1802A}
}

@ARTICLE{2025PhRvL.134o1001Z,
	author = {{Zimmermann}, Tim and {Alvey}, James and {Marsh}, David J.~E. and {Fairbairn}, Malcolm and {Read}, Justin I.},
	title = "{Dwarf Galaxies Imply Dark Matter is Heavier than 2.2{\texttimes}10-21  eV}",
	journal = {\prl},
	year = 2025,
	month = apr,
	volume = {134},
	number = {15},
	eid = {151001},
	pages = {151001},
	url = {https://doi.org/10.1103/PhysRevLett.134.151001},
	doi = {10.1103/PhysRevLett.134.151001},
	archivePrefix = {arXiv},
	eprint = {2405.20374},
	primaryClass = {astro-ph.CO},
	adsurl = {https://ui.adsabs.harvard.edu/abs/2025PhRvL.134o1001Z}
}

@ARTICLE{2025arXiv251208065A,
	author = {{Akerib}, D.~S. and {Al Musalhi}, A.~K. and {Alder}, F. and {Almquist}, B.~J. and {Amarasinghe}, C.~S. and {Ames}, A. and {Anderson}, T.~J. and {Angelides}, N. and {Ara{\'u}jo}, H.~M. and {Armstrong}, J.~E. and {Arthurs}, M. and {Baker}, A. and {Balashov}, S. and {Bang}, J. and {Bargemann}, J.~W. and {Barillier}, E.~E. and {Bauer}, D. and {Beattie}, K. and {Bhatti}, A. and {Biesiadzinski}, T.~P. and {Birch}, H.~J. and {Bishop}, E. and {Blockinger}, G.~M. and {Brew}, C.~A.~J. and {Br{\'a}s}, P. and {Burdin}, S. and {Carmona-Benitez}, M.~C. and {Carter}, M. and {Chawla}, A. and {Chen}, H. and {Chin}, Y.~T. and {Chott}, N.~I. and {Contreras}, S. and {Converse}, M.~V. and {Coronel}, R. and {Cottle}, A. and {Cox}, G. and {Curran}, D. and {Dahl}, C.~E. and {Darlington}, I. and {Dave}, S. and {David}, A. and {Delgaudio}, J. and {Dey}, S. and {de Viveiros}, L. and {Di Felice}, L. and {Ding}, C. and {Dobson}, J.~E.~Y. and {Druszkiewicz}, E. and {Dubey}, S. and {Dunbar}, C.~L. and {Eriksen}, S.~R. and {Fayer}, S. and {Fearon}, N.~M. and {Fieldhouse}, N. and {Fiorucci}, S. and {Flaecher}, H. and {Fraser}, E.~D. and {Fruth}, T.~M.~A. and {Gaemers}, P.~W. and {Gaitskell}, R.~J. and {Geffre}, A. and {Genovesi}, J. and {Ghag}, C. and {Ghamsari}, J. and {Ghosh}, A. and {Ghosh}, S. and {Gibbons}, R. and {Gokhale}, S. and {Green}, J. and {van der Grinten}, M.~G.~D. and {Haiston}, J.~J. and {Hall}, C.~R. and {Hall}, T. and {Hampp}, R. H and {Haselschwardt}, S.~J. and {Hernandez}, M.~A. and {Hertel}, S.~A. and {Homenides}, G.~J. and {Horn}, M. and {Huang}, D.~Q. and {Hunt}, D. and {Jacquet}, E. and {James}, R.~S. and {Jenkins}, K. and {Kaboth}, A.~C. and {Kamaha}, A.~C. and {Kannichankandy}, M.~K. and {Khaitan}, D. and {Khazov}, A. and {Kim}, J. and {Kim}, Y.~D. and {Kodroff}, D. and {Korolkova}, E.~V. and {Kraus}, H. and {Kravitz}, S. and {Kreczko}, L. and {Kudryavtsev}, V.~A. and {Lawes}, C. and {Leonard}, D.~S. and {Lesko}, K.~T. and {Levy}, C. and {Lin}, J. and {Lindote}, A. and {Lippincott}, W.~H. and {Long}, J. and {Lopes}, M.~I. and {Lorenzon}, W. and {Lu}, C. and {Luitz}, S. and {Ma}, W. and {Mahajan}, V. and {Majewski}, P.~A. and {Manalaysay}, A. and {Mannino}, R.~L. and {Matheson}, R.~J. and {Maupin}, C. and {McCarthy}, M.~E. and {McKinsey}, D.~N. and {McLaughlin}, J. and {McLaughlin}, J.~B. and {McMonigle}, R. and {Mitra}, B. and {Mizrachi}, E. and {Monzani}, M.~E. and {Mor{\r{a}}}, K. and {Morrison}, E. and {Mount}, B.~J. and {Murdy}, M. and {Murphy}, A. St. J. and {Nelson}, H.~N. and {Neves}, F. and {Nguyen}, A. and {O'Brien}, C.~L. and {O'Shea}, F.~H. and {Olcina}, I. and {Oliver-Mallory}, K.~C. and {Orpwood}, J. and {Y Oyulmaz}, K. and {Palladino}, K.~J. and {Pannifer}, N.~J. and {Parveen}, N. and {Patton}, S.~J. and {Penning}, B. and {Pereira}, G. and {Perry}, E. and {Pershing}, T. and {Piepke}, A. and {Poudel}, S.~S. and {Qie}, Y. and {Reichenbacher}, J. and {Rhyne}, C.~A. and {Rischbieter}, G.~R.~C. and {Ritchey}, E. and {Riyat}, H.~S. and {Rosero}, R. and {Rowe}, N.~J. and {Rushton}, T. and {Rynders}, D. and {Salt{\~a}o}, S. and {Santone}, D. and {Sargeant}, I. and {Sazzad}, A.~B.~M.~R. and {Schnee}, R.~W. and {Sehr}, G. and {Shafer}, B. and {Shaw}, S. and {Sherman}, W. and {Shi}, K. and {Shutt}, T. and {Silva}, C. and {Sinev}, G. and {Siniscalco}, J. and {Slivar}, A.~M. and {Smith}, R. and {Solovov}, V.~N. and {Sorensen}, P. and {Soria}, J. and {Sumner}, T.~J. and {Swain}, A. and {Szydagis}, M. and {Tiedt}, D.~R. and {Timalsina}, M. and {Tovey}, D.~R. and {Tranter}, J. and {Trask}, M. and {Trengove}, K. and {Tripathi}, M. and {Us{\'o}n}, A. and {Vaitkus}, A.~C. and {Valentino}, O. and {Velan}, V. and {Wang}, A. and {Wang}, J.~J. and {Wang}, Y. and {Weeldreyer}, L. and {Whitis}, T.~J. and {Wild}, K. and {Williams}, M. and {Winnicki}, J.},
	title = "{Searches for Light Dark Matter and Evidence of Coherent Elastic Neutrino-Nucleus Scattering of Solar Neutrinos with the LUX-ZEPLIN (LZ) Experiment}",
	journal = {arXiv e-prints},
	year = 2025,
	month = dec,
	eid = {arXiv:2512.08065},
	pages = {arXiv:2512.08065},
	url = {https://doi.org/10.48550/arXiv.2512.08065},
	doi = {10.48550/arXiv.2512.08065},
	archivePrefix = {arXiv},
	eprint = {2512.08065},
	primaryClass = {hep-ex},
	adsurl = {https://ui.adsabs.harvard.edu/abs/2025arXiv251208065A}
}

@ARTICLE{2025arXiv251114593A,
	author = {{Abusleme}, Angel and {Adam}, Thomas and {Adamowicz}, Kai and {Adey}, David and {Ahmad}, Shakeel and {Ahmed}, Rizwan and {Ahola}, Timo and {Aiello}, Sebastiano and {An}, Fengpeng and {An}, Guangpeng and {Andreopoulos}, Costas and {Andronico}, Giuseppe and {Athayde Marcondes de Andr{\'e}}, Jo{\~a}o Pedro and {Anfimov}, Nikolay and {Antonelli}, Vito and {Antoshkina}, Tatiana and {Asavapibhop}, Burin and {Auguste}, Didier and {Buizza Avanzini}, Margherita and {Babic}, Andrej and {Bai}, Jingzhi and {Bai}, Weidong and {Balashov}, Nikita and {Barbera}, Roberto and {Barresi}, Andrea and {Basilico}, Davide and {Baussan}, Eric and {Bellantonio}, Beatrice and {Bellato}, Marco and {Beney}, JeanLuc and {Beretta}, Marco and {Bergnoli}, Antonio and {Bernieri}, Enrico and {Bessonov}, Nikita and {Biar{\'e}}, David and {Bick}, Daniel and {Bieger}, Lukas and {Biktemerova}, Svetlana and {Birkenfeld}, Thilo and {Blum}, David and {Blyth}, Simon and {Boarin}, Sara and {Boehles}, Manuel and {Bolshakova}, Anastasia and {Bongrand}, Mathieu and {Bonhomme}, Aur{\'e}lie and {Bordereau}, Cl{\'e}ment and {Borghesi}, Matteo and {Brigatti}, Augusto and {Brugiere}, Timothee and {Brugnera}, Riccardo and {Bruno}, Riccardo and {Buchholz}, Jonas and {Budano}, Antonio and {Buesken}, Max and {Buscemi}, Mario and {Bussino}, Severino and {Busto}, Jose and {Butorov}, Ilya and {B{\"u}chner}, Marcel and {Cabrera}, Anatael and {Caccianiga}, Barbara and {Cai}, Boshuai and {Cai}, Hao and {Cai}, Xiao and {Cai}, Yanke and {Cai}, Yi-zhou and {Cai}, Zhiyan and {Callier}, St{\'e}phane and {Calvez}, Steven and {Cammi}, Antonio and {Campeny}, Agustin and {Cai}, Dechang and {Cao}, Chuanya and {Cao}, Dewen and {Cao}, Guofu and {Cao}, Jun and {Cao}, Yaoqi and {Caruso}, Rossella and {Caslini}, Aurelio and {Cerna}, C{\'e}dric and {Cerrone}, Vanessa and {Cesini}, Daniele and {Chan}, Chi and {Chang}, Jinfan and {Chang}, Yun and {Charavet}, Milo and {Chariss{\'e}}, Tim and {Chatrabhuti}, Auttakit and {Chen}, Chao and {Chen}, Guoming and {Chen}, Haitao and {Chen}, Haotian and {Chen}, Jiahui and {Chen}, Jian and {Chen}, Jing and {Chen}, Junyou and {Chen}, Lihao and {Chen}, Mali and {Chen}, Mingming and {Chen}, Pingping and {Chen}, Po-An and {Chen}, Quanyou and {Chen}, Shaomin and {Chen}, Shenjian and {Chen}, Shi and {Chen}, Shiqiang and {Chen}, Sisi and {Chen}, Xin and {Chen}, Xuan and {Chen}, Xurong and {Chen}, Yi-Wen and {Chen}, Yiming and {Chen}, Yixue and {Chen}, Yu and {Chen}, Ze and {Chen}, Zelin and {Chen}, Zhang and {Chen}, Zhangming and {Chen}, Zhiyuan and {Chen}, Zhongchang and {Chen}, Zikang and {Cheng}, Brian and {Cheng}, Jie and {Cheng}, Yaping and {Cheng}, Yu Chin and {Cheng}, Zhaokan and {Chepurnov}, Alexander and {Chetverikov}, Alexey and {Chiesa}, Davide and {Chimenti}, Pietro and {Chin}, Yen-Ting and {Chiu}, Pin-Jung and {Chou}, Po-Lin and {Chu}, Ziliang and {Chukanov}, Artem and {Singh Chundawat}, Neetu Raj and {Chuvashova}, Anna and {Claverie}, G{\'e}rard and {Clementi}, Catia and {Clerbaux}, Barbara and {Coletta}, Claudio and {Colomer Molla}, Marta and {Dal Corso}, Flavio and {Corti}, Daniele and {Costa}, Salvatore and {Csakli}, Simon and {Cui}, Chenyang and {Cui}, Shanshan and {Vincenzo D'Auria}, Lorenzo and {Dalager}, Olivia and {Datta}, Jaydeep and {Delgadillo Franco}, Luis and {Deng}, Jiawei and {Deng}, Zhi and {Deng}, Ziyan and {Depnering}, Wilfried and {Didenko}, Hanna and {Ding}, Xiaoyu and {Ding}, Xuefeng and {Ding}, Yayun and {Dirgantara}, Bayu and {Dittrich}, Carsten and {Dmitrievsky}, Sergey and {Doerflinger}, David and {Dohnal}, Tadeas and {Dolgareva}, Maria and {Dolzhikov}, Dmitry and {Dong}, Chuanshi and {Dong}, Haojie and {Dong}, Jianmeng and {Dong}, Lan and {Dornic}, Damien and {Doroshkevich}, Evgeny and {Dou}, Wei and {Dracos}, Marcos and {Drapier}, Olivier and {Drohmann}, Tobias and {Druillole}, Fr{\'e}d{\'e}ric and {Du}, Ran and {Du}, Shuxian and {Duan}, Yujie and {Dugas}, Katherine and {Dusini}, Stefano and {Duyang}, Hongyue and {Dvorak}, Martin and {Eck}, Jessica and {Enqvist}, Timo and {Enzmann}, Heike and {Fabbri}, Andrea and {Fahrendholz}, Ulrike and {Fajt}, Lukas and {Fan}, Donghua and {Fan}, Lei and {Fan}, Liangqianjin and {Fang}, Can and {Fang}, Jian and {Fang}, Wenxing and {Fargetta}, Marco and {Stanescu Farilla}, Elia},
	title = "{First measurement of reactor neutrino oscillations at JUNO}",
	journal = {arXiv e-prints},
	year = 2025,
	month = nov,
	eid = {arXiv:2511.14593},
	pages = {arXiv:2511.14593},
	url = {https://doi.org/10.48550/arXiv.2511.14593},
	doi = {10.48550/arXiv.2511.14593},
	archivePrefix = {arXiv},
	eprint = {2511.14593},
	primaryClass = {hep-ex},
	adsurl = {https://ui.adsabs.harvard.edu/abs/2025arXiv251114593A}
}

@ARTICLE{2025EPJC...85.1192X,
	author = {{XLZD Collaboration} and {Aalbers}, J. and {Abe}, K. and {Adrover}, M. and {Ahmed Maouloud}, S. and {Akerib}, D.~S. and {Al Musalhi}, A.~K. and {Alder}, F. and {Althueser}, L. and {Amaral}, D.~W.~P. and {Amarasinghe}, C.~S. and {Ames}, A. and {Andrieu}, B. and {Angelides}, N. and {Angelino}, E. and {Antunovic}, B. and {Aprile}, E. and {Ara{\'u}jo}, H.~M. and {Armstrong}, J.~E. and {Arthurs}, M. and {Babicz}, M. and {Baker}, A. and {Balzer}, M. and {Bang}, J. and {Barberio}, E. and {Bargemann}, J.~W. and {Barillier}, E. and {Basharina-Freshville}, A. and {Baudis}, L. and {Bauer}, D. and {Bazyk}, M. and {Beattie}, K. and {Beaupere}, N. and {Bell}, N.~F. and {Bellagamba}, L. and {Benson}, T. and {Bhatti}, A. and {Biesiadzinski}, T.~P. and {Biondi}, R. and {Biondi}, Y. and {Birch}, H.~J. and {Bishop}, E. and {Bismark}, A. and {Boehm}, C. and {Boese}, K. and {Bolotnikov}, A. and {Br{\'a}s}, P. and {Braun}, R. and {Breskin}, A. and {Brew}, C.~A.~J. and {Brommer}, S. and {Brown}, A. and {Bruni}, G. and {Budnik}, R. and {Burdin}, S. and {Cai}, C. and {Capelli}, C. and {Carini}, G. and {Carmona-Benitez}, M.~C. and {Carter}, M. and {Chauvin}, A. and {Chawla}, A. and {Chen}, H. and {Cherwinka}, J.~J. and {Chin}, Y.~T. and {Chott}, N.~I. and {Chavez}, A.~P. Cimental and {Clark}, K. and {Colijn}, A.~P. and {Colling}, D.~J. and {Conrad}, J. and {Converse}, M.~V. and {Cooper}, L.~J. and {Coronel}, R. and {Costanzo}, D. and {Cottle}, A. and {Cox}, G. and {Cuenca-Garc{\'\i}a}, J.~J. and {Curran}, D. and {Cussans}, D. and {D'Andrea}, V. and {Daniel Garcia}, L.~C. and {Darlington}, I. and {Dave}, S. and {David}, A. and {Davies}, G.~J. and {Decowski}, M.~P. and {Deisting}, A. and {Delgaudio}, J. and {Dey}, S. and {Di Donato}, C. and {Di Felice}, L. and {Di Gangi}, P. and {Diglio}, S. and {Ding}, C. and {Dobson}, J.~E.~Y. and {Doerenkamp}, M. and {Drexlin}, G. and {Druszkiewicz}, E. and {Dunbar}, C.~L. and {Eitel}, K. and {Elykov}, A. and {Engel}, R. and {Eriksen}, S.~R. and {Fayer}, S. and {Fearon}, N.~M. and {Ferella}, A.~D. and {Ferrari}, C. and {Fieldhouse}, N. and {Fischer}, H. and {Flaecher}, H. and {Flehmke}, T. and {Flierman}, M. and {Fraser}, E.~D. and {Fruth}, T.~M.~A. and {Fujikawa}, K. and {Fulgione}, W. and {Fuselli}, C. and {Gaemers}, P. and {Gaior}, R. and {Gaitskell}, R.~J. and {Gallice}, N. and {Galloway}, M. and {Gao}, F. and {Garroum}, N. and {Geffre}, A. and {Genovesi}, J. and {Ghag}, C. and {Ghosh}, S. and {Giacomobono}, R. and {Gibbons}, R. and {Girard}, F. and {Glade-Beucke}, R. and {Gl{\"u}ck}, F. and {Gokhale}, S. and {Grandi}, L. and {Green}, J. and {Grigat}, J. and {van der Grinten}, M.~G.~D. and {Gr{\"o}{\ss}le}, R. and {Guan}, H. and {Guida}, M. and {Gyorgy}, P. and {Haiston}, J.~J. and {Hall}, C.~R. and {Hall}, T. and {Hammann}, R. and {Hannen}, V. and {Hansmann-Menzemer}, S. and {Hargittai}, N. and {Hartigan-O'Connor}, E. and {Haselschwardt}, S.~J. and {Hernandez}, M. and {Hertel}, S.~A. and {Higuera}, A. and {Hils}, C. and {Hiraoka}, K. and {Hoetzsch}, L. and {Hoferichter}, M. and {Homenides}, G.~J. and {Hood}, N.~F. and {Horn}, M. and {Huang}, D.~Q. and {Hughes}, S. and {Hunt}, D. and {Iacovacci}, M. and {Itow}, Y. and {Jacquet}, E. and {Jakob}, J. and {James}, R.~S. and {Joerg}, F. and {Jones}, S. and {Kaboth}, A.~C. and {Kahlert}, F. and {Kamaha}, A.~C. and {Kaminaga}, Y. and {Kara}, M. and {Kavrigin}, P. and {Kazama}, S. and {Keller}, M. and {Kemp-Russell}, P. and {Khaitan}, D. and {Kharbanda}, P. and {Kilminster}, B. and {Kim}, J. and {Kirk}, R. and {Kleifges}, M. and {Klute}, M. and {Kobayashi}, M. and {Kodroff}, D. and {Koke}, D. and {Kopec}, A. and {Korolkova}, E. v. and {Kraus}, H. and {Kravitz}, S. and {Kreczko}, L. and {von Krosigk}, B. and {Kudryavtsev}, V.~A. and {Kuger}, F. and {Kurita}, N.},
	title = "{The XLZD Design Book: towards the next-generation liquid xenon observatory for dark matter and neutrino physics}",
	journal = {European Physical Journal C},
	year = 2025,
	month = oct,
	volume = {85},
	number = {10},
	eid = {1192},
	pages = {1192},
	url = {https://doi.org/10.1140/epjc/s10052-025-14810-w},
	doi = {10.1140/epjc/s10052-025-14810-w},
	archivePrefix = {arXiv},
	eprint = {2410.17137},
	primaryClass = {hep-ex},
	adsurl = {https://ui.adsabs.harvard.edu/abs/2025EPJC...85.1192X}
}

@ARTICLE{2023JHEP...07..071A,
	author = {{Amaral}, Dorian and {Cerde{\~n}o}, David and {Cheek}, Andrew and {Foldenauer}, Patrick},
	title = "{A direct detection view of the neutrino NSI landscape}",
	journal = {Journal of High Energy Physics},
	year = 2023,
	month = jul,
	volume = {2023},
	number = {7},
	eid = {71},
	pages = {71},
	url = {https://doi.org/10.1007/JHEP07(2023)071},
	doi = {10.1007/JHEP07(2023)071},
	archivePrefix = {arXiv},
	eprint = {2302.12846},
	primaryClass = {hep-ph},
	adsurl = {https://ui.adsabs.harvard.edu/abs/2023JHEP...07..071A}
}

@ARTICLE{2021EPJC...81.1015A,
	author = {{Agrawal}, P. and {Bauer}, M. and {Beacham}, J. and {Berlin}, A. and {Boyarsky}, A. and {Cebrian}, S. and {Cid-Vidal}, X. and {d'Enterria}, D. and {De Roeck}, A. and {Drewes}, M. and {Echenard}, B. and {Giannotti}, M. and {Giudice}, G.~F. and {Gninenko}, S. and {Gori}, S. and {Goudzovski}, E. and {Heeck}, J. and {Hernandez}, P. and {Hostert}, M. and {Irastorza}, I.~G. and {Izmaylov}, A. and {Jaeckel}, J. and {Kahlhoefer}, F. and {Knapen}, S. and {Krnjaic}, G. and {Lanfranchi}, G. and {Monroe}, J. and {Outschoorn}, V.~I. Martinez and {Lopez-Pavon}, J. and {Pascoli}, S. and {Pospelov}, M. and {Redigolo}, D. and {Ringwald}, A. and {Ruchayskiy}, O. and {Ruderman}, J. and {Russell}, H. and {Salfeld-Nebgen}, J. and {Schuster}, P. and {Shaposhnikov}, M. and {Shchutska}, L. and {Shelton}, J. and {Soreq}, Y. and {Stadnik}, Y. and {Swallow}, J. and {Tobioka}, K. and {Tsai}, Y. -D.},
	title = "{Feebly-interacting particles: FIPs 2020 workshop report}",
	journal = {European Physical Journal C},
	year = 2021,
	month = nov,
	volume = {81},
	number = {11},
	eid = {1015},
	pages = {1015},
	url = {https://doi.org/10.1140/epjc/s10052-021-09703-7},
	doi = {10.1140/epjc/s10052-021-09703-7},
	archivePrefix = {arXiv},
	eprint = {2102.12143},
	primaryClass = {hep-ph},
	adsurl = {https://ui.adsabs.harvard.edu/abs/2021EPJC...81.1015A}
}

@ARTICLE{2022PrPNP.12403947A,
	author = {{Athar}, M. Sajjad and {Barwick}, Steven W. and {Brunner}, Thomas and {Cao}, Jun and {Danilov}, Mikhail and {Inoue}, Kunio and {Kajita}, Takaaki and {Kowalski}, Marek and {Lindner}, Manfred and {Long}, Kenneth R. and {Palanque-Delabrouille}, Nathalie and {Rodejohann}, Werner and {Schellman}, Heidi and {Scholberg}, Kate and {Seo}, Seon-Hee and {Smith}, Nigel J.~T. and {Winter}, Walter and {Zeller}, Geralyn P. and {Funchal}, Renata Zukanovich},
	title = "{Status and perspectives of neutrino physics}",
	journal = {Progress in Particle and Nuclear Physics},
	year = 2022,
	month = may,
	volume = {124},
	eid = {103947},
	pages = {103947},
	url = {https://doi.org/10.1016/j.ppnp.2022.103947},
	doi = {10.1016/j.ppnp.2022.103947},
	archivePrefix = {arXiv},
	eprint = {2111.07586},
	primaryClass = {hep-ph},
	adsurl = {https://ui.adsabs.harvard.edu/abs/2022PrPNP.12403947A}
}

@ARTICLE{2013PhRvC..88b5501A,
	author = {{Aharmim}, B. and {Ahmed}, S.~N. and {Anthony}, A.~E. and {Barros}, N. and {Beier}, E.~W. and {Bellerive}, A. and {Beltran}, B. and {Bergevin}, M. and {Biller}, S.~D. and {Boudjemline}, K. and {Boulay}, M.~G. and {Cai}, B. and {Chan}, Y.~D. and {Chauhan}, D. and {Chen}, M. and {Cleveland}, B.~T. and {Cox}, G.~A. and {Dai}, X. and {Deng}, H. and {Detwiler}, J.~A. and {DiMarco}, M. and {Doe}, P.~J. and {Doucas}, G. and {Drouin}, P. -L. and {Duncan}, F.~A. and {Dunford}, M. and {Earle}, E.~D. and {Elliott}, S.~R. and {Evans}, H.~C. and {Ewan}, G.~T. and {Farine}, J. and {Fergani}, H. and {Fleurot}, F. and {Ford}, R.~J. and {Formaggio}, J.~A. and {Gagnon}, N. and {Goon}, J. TM. and {Graham}, K. and {Guillian}, E. and {Habib}, S. and {Hahn}, R.~L. and {Hallin}, A.~L. and {Hallman}, E.~D. and {Harvey}, P.~J. and {Hazama}, R. and {Heintzelman}, W.~J. and {Heise}, J. and {Helmer}, R.~L. and {Hime}, A. and {Howard}, C. and {Huang}, M. and {Jagam}, P. and {Jamieson}, B. and {Jelley}, N.~A. and {Jerkins}, M. and {Keeter}, K.~J. and {Klein}, J.~R. and {Kormos}, L.~L. and {Kos}, M. and {Kraus}, C. and {Krauss}, C.~B. and {Kruger}, A. and {Kutter}, T. and {Kyba}, C.~C.~M. and {Lange}, R. and {Law}, J. and {Lawson}, I.~T. and {Lesko}, K.~T. and {Leslie}, J.~R. and {Loach}, J.~C. and {MacLellan}, R. and {Majerus}, S. and {Mak}, H.~B. and {Maneira}, J. and {Martin}, R. and {McCauley}, N. and {McDonald}, A.~B. and {McGee}, S.~R. and {Miller}, M.~L. and {Monreal}, B. and {Monroe}, J. and {Nickel}, B.~G. and {Noble}, A.~J. and {O'Keeffe}, H.~M. and {Oblath}, N.~S. and {Ollerhead}, R.~W. and {Orebi Gann}, G.~D. and {Oser}, S.~M. and {Ott}, R.~A. and {Peeters}, S.~J.~M. and {Poon}, A.~W.~P. and {Prior}, G. and {Reitzner}, S.~D. and {Rielage}, K. and {Robertson}, B.~C. and {Robertson}, R.~G.~H. and {Rosten}, R.~C. and {Schwendener}, M.~H. and {Secrest}, J.~A. and {Seibert}, S.~R. and {Simard}, O. and {Simpson}, J.~J. and {Skensved}, P. and {Sonley}, T.~J. and {Stonehill}, L.~C. and {Te{\v{s}}i{\'c}}, G. and {Tolich}, N. and {Tsui}, T. and {Van Berg}, R. and {VanDevender}, B.~A. and {Virtue}, C.~J. and {Wan Chan Tseung}, H. and {Wark}, D.~L. and {Watson}, P.~J.~S. and {Wendland}, J. and {West}, N. and {Wilkerson}, J.~F. and {Wilson}, J.~R. and {Wouters}, J.~M. and {Wright}, A. and {Yeh}, M. and {Zhang}, F. and {Zuber}, K.},
	title = "{Combined analysis of all three phases of solar neutrino data from the Sudbury Neutrino Observatory}",
	journal = {\prc},
	year = 2013,
	month = aug,
	volume = {88},
	number = {2},
	eid = {025501},
	pages = {025501},
	url = {https://doi.org/10.1103/PhysRevC.88.025501},
	doi = {10.1103/PhysRevC.88.025501},
	archivePrefix = {arXiv},
	eprint = {1109.0763},
	primaryClass = {nucl-ex},
	adsurl = {https://ui.adsabs.harvard.edu/abs/2013PhRvC..88b5501A}
}

@ARTICLE{2011PhRvC..84c5804A,
	author = {{Abe}, S. and {Furuno}, K. and {Gando}, A. and {Gando}, Y. and {Ichimura}, K. and {Ikeda}, H. and {Inoue}, K. and {Kibe}, Y. and {Kimura}, W. and {Kishimoto}, Y. and {Koga}, M. and {Minekawa}, Y. and {Mitsui}, T. and {Morikawa}, T. and {Nagai}, N. and {Nakajima}, K. and {Nakamura}, K. and {Nakamura}, M. and {Narita}, K. and {Shimizu}, I. and {Shimizu}, Y. and {Shirai}, J. and {Suekane}, F. and {Suzuki}, A. and {Takahashi}, H. and {Takahashi}, N. and {Takemoto}, Y. and {Tamae}, K. and {Watanabe}, H. and {Xu}, B.~D. and {Yabumoto}, H. and {Yonezawa}, E. and {Yoshida}, H. and {Yoshida}, S. and {Enomoto}, S. and {Kozlov}, A. and {Murayama}, H. and {Grant}, C. and {Keefer}, G. and {McKee}, D. and {Piepke}, A. and {Banks}, T.~I. and {Bloxham}, T. and {Detwiler}, J.~A. and {Freedman}, S.~J. and {Fujikawa}, B.~K. and {Han}, K. and {Kadel}, R. and {O'Donnell}, T. and {Steiner}, H.~M. and {Winslow}, L.~A. and {Dwyer}, D.~A. and {Mauger}, C. and {McKeown}, R.~D. and {Zhang}, C. and {Berger}, B.~E. and {Lane}, C.~E. and {Maricic}, J. and {Miletic}, T. and {Batygov}, M. and {Learned}, J.~G. and {Matsuno}, S. and {Pakvasa}, S. and {Sakai}, M. and {Horton-Smith}, G.~A. and {Tang}, A. and {Downum}, K.~E. and {Gratta}, G. and {Tolich}, K. and {Efremenko}, Y. and {Kamyshkov}, Y. and {Perevozchikov}, O. and {Karwowski}, H.~J. and {Markoff}, D.~M. and {Tornow}, W. and {Heeger}, K.~M. and {Piquemal}, F. and {Ricol}, J. -S. and {Decowski}, M.~P.},
	title = "{Measurement of the $^{8}$B solar neutrino flux with the KamLAND liquid scintillator detector}",
	journal = {\prc},
	year = 2011,
	month = sep,
	volume = {84},
	number = {3},
	eid = {035804},
	pages = {035804},
	url = {https://doi.org/10.1103/PhysRevC.84.035804},
	doi = {10.1103/PhysRevC.84.035804},
	archivePrefix = {arXiv},
	eprint = {1106.0861},
	primaryClass = {hep-ex},
	adsurl = {https://ui.adsabs.harvard.edu/abs/2011PhRvC..84c5804A}
}

@ARTICLE{2016PhRvD..94e2010A,
	author = {{Abe}, K. and {Haga}, Y. and {Hayato}, Y. and {Ikeda}, M. and {Iyogi}, K. and {Kameda}, J. and {Kishimoto}, Y. and {Marti}, Ll. and {Miura}, M. and {Moriyama}, S. and {Nakahata}, M. and {Nakajima}, T. and {Nakayama}, S. and {Orii}, A. and {Sekiya}, H. and {Shiozawa}, M. and {Sonoda}, Y. and {Takeda}, A. and {Tanaka}, H. and {Takenaga}, Y. and {Tasaka}, S. and {Tomura}, T. and {Ueno}, K. and {Yokozawa}, T. and {Akutsu}, R. and {Irvine}, T. and {Kaji}, H. and {Kajita}, T. and {Kametani}, I. and {Kaneyuki}, K. and {Lee}, K.~P. and {Nishimura}, Y. and {McLachlan}, T. and {Okumura}, K. and {Richard}, E. and {Labarga}, L. and {Fernandez}, P. and {Blaszczyk}, F. d. M. and {Gustafson}, J. and {Kachulis}, C. and {Kearns}, E. and {Raaf}, J.~L. and {Stone}, J.~L. and {Sulak}, L.~R. and {Berkman}, S. and {Tobayama}, S. and {Goldhaber}, M. and {Bays}, K. and {Carminati}, G. and {Griskevich}, N.~J. and {Kropp}, W.~R. and {Mine}, S. and {Renshaw}, A. and {Smy}, M.~B. and {Sobel}, H.~W. and {Takhistov}, V. and {Weatherly}, P. and {Ganezer}, K.~S. and {Hartfiel}, B.~L. and {Hill}, J. and {Keig}, W.~E. and {Hong}, N. and {Kim}, J.~Y. and {Lim}, I.~T. and {Park}, R.~G. and {Akiri}, T. and {Albert}, J.~B. and {Himmel}, A. and {Li}, Z. and {O'Sullivan}, E. and {Scholberg}, K. and {Walter}, C.~W. and {Wongjirad}, T. and {Ishizuka}, T. and {Nakamura}, T. and {Jang}, J.~S. and {Choi}, K. and {Learned}, J.~G. and {Matsuno}, S. and {Smith}, S.~N. and {Friend}, M. and {Hasegawa}, T. and {Ishida}, T. and {Ishii}, T. and {Kobayashi}, T. and {Nakadaira}, T. and {Nakamura}, K. and {Nishikawa}, K. and {Oyama}, Y. and {Sakashita}, K. and {Sekiguchi}, T. and {Tsukamoto}, T. and {Nakano}, Y. and {Suzuki}, A.~T. and {Takeuchi}, Y. and {Yano}, T. and {Cao}, S.~V. and {Hayashino}, T. and {Hiraki}, T. and {Hirota}, S. and {Huang}, K. and {Ieki}, K. and {Jiang}, M. and {Kikawa}, T. and {Minamino}, A. and {Murakami}, A. and {Nakaya}, T. and {Patel}, N.~D. and {Suzuki}, K. and {Takahashi}, S. and {Wendell}, R.~A. and {Fukuda}, Y. and {Itow}, Y. and {Mitsuka}, G. and {Muto}, F. and {Suzuki}, T. and {Mijakowski}, P. and {Frankiewicz}, K. and {Hignight}, J. and {Imber}, J. and {Jung}, C.~K. and {Li}, X. and {Palomino}, J.~L. and {Santucci}, G. and {Taylor}, I. and {Vilela}, C. and {Wilking}, M.~J. and {Yanagisawa}, C. and {Fukuda}, D. and {Ishino}, H. and {Kayano}, T. and {Kibayashi}, A. and {Koshio}, Y. and {Mori}, T. and {Sakuda}, M. and {Takeuchi}, J. and {Yamaguchi}, R. and {Kuno}, Y. and {Tacik}, R. and {Kim}, S.~B. and {Okazawa}, H. and {Choi}, Y. and {Ito}, K. and {Nishijima}, K. and {Koshiba}, M. and {Totsuka}, Y. and {Suda}, Y. and {Yokoyama}, M. and {Bronner}, C. and {Calland}, R.~G. and {Hartz}, M. and {Martens}, K. and {Obayashi}, Y. and {Suzuki}, Y. and {Vagins}, M.~R. and {Nantais}, C.~M. and {Martin}, J.~F. and {de Perio}, P. and {Tanaka}, H.~A. and {Konaka}, A. and {Chen}, S. and {Sui}, H. and {Wan}, L. and {Yang}, Z. and {Zhang}, H. and {Zhang}, Y. and {Connolly}, K. and {Dziomba}, M. and {Wilkes}, R.~J. and {Super-Kamiokande Collaboration}},
	title = "{Solar neutrino measurements in Super-Kamiokande-IV}",
	journal = {\prd},
	year = 2016,
	month = sep,
	volume = {94},
	number = {5},
	eid = {052010},
	pages = {052010},
	url = {https://doi.org/10.1103/PhysRevD.94.052010},
	doi = {10.1103/PhysRevD.94.052010},
	archivePrefix = {arXiv},
	eprint = {1606.07538},
	primaryClass = {hep-ex},
	adsurl = {https://ui.adsabs.harvard.edu/abs/2016PhRvD..94e2010A}
}

@ARTICLE{2021PrPNP.11903865A,
	author = {{Arbey}, A. and {Mahmoudi}, F.},
	title = "{Dark matter and the early Universe: A review}",
	journal = {Progress in Particle and Nuclear Physics},
	year = 2021,
	month = jul,
	volume = {119},
	eid = {103865},
	pages = {103865},
	url = {https://doi.org/10.1016/j.ppnp.2021.103865},
	doi = {10.1016/j.ppnp.2021.103865},
	archivePrefix = {arXiv},
	eprint = {2104.11488},
	primaryClass = {hep-ph},
	adsurl = {https://ui.adsabs.harvard.edu/abs/2021PrPNP.11903865A}
}

@ARTICLE{2019PhRvD.100h2004A,
	author = {{Agostini}, M. and {Altenm{\"u}ller}, K. and {Appel}, S. and {Atroshchenko}, V. and {Bagdasarian}, Z. and {Basilico}, D. and {Bellini}, G. and {Benziger}, J. and {Bonfini}, G. and {Bravo}, D. and {Caccianiga}, B. and {Calaprice}, F. and {Caminata}, A. and {Cappelli}, L. and {Caprioli}, S. and {Carlini}, M. and {Cavalcante}, P. and {Cavanna}, F. and {Chepurnov}, A. and {Choi}, K. and {Collica}, L. and {D'Angelo}, D. and {Davini}, S. and {Derbin}, A. and {Ding}, X.~F. and {Di Ludovico}, A. and {Di Noto}, L. and {Drachnev}, I. and {Fomenko}, K. and {Formozov}, A. and {Franco}, D. and {Gabriele}, F. and {Galbiati}, C. and {Gschwender}, M. and {Ghiano}, C. and {Giammarchi}, M. and {Goretti}, A. and {Gromov}, M. and {Guffanti}, D. and {Houdy}, T. and {Hungerford}, E. and {Ianni}, Aldo and {Ianni}, Andrea and {Jany}, A. and {Jeschke}, D. and {Kumaran}, S. and {Kobychev}, V. and {Korga}, G. and {Lachenmaier}, T. and {Laubenstein}, M. and {Litvinovich}, E. and {Lombardi}, P. and {Ludhova}, L. and {Lukyanchenko}, G. and {Lukyanchenko}, L. and {Machulin}, I. and {Manuzio}, G. and {Marcocci}, S. and {Maricic}, J. and {Martyn}, J. and {Meroni}, E. and {Meyer}, M. and {Miramonti}, L. and {Misiaszek}, M. and {Muratova}, V. and {Neumair}, B. and {Nieslony}, M. and {Oberauer}, L. and {Orekhov}, V. and {Ortica}, F. and {Pallavicini}, M. and {Papp}, L. and {Penek}, {\"O}. and {Pietrofaccia}, L. and {Pilipenko}, N. and {Pocar}, A. and {Porcelli}, A. and {Raikov}, G. and {Ranucci}, G. and {Razeto}, A. and {Re}, A. and {Redchuk}, M. and {Romani}, A. and {Rossi}, N. and {Rottenanger}, S. and {Sch{\"o}nert}, S. and {Semenov}, D. and {Skorokhvatov}, M. and {Smirnov}, O. and {Sotnikov}, A. and {Stokes}, L.~F.~F. and {Suvorov}, Y. and {Tartaglia}, R. and {Testera}, G. and {Thurn}, J. and {Unzhakov}, E. and {Villante}, F. and {Vishneva}, A. and {Vogelaar}, R.~B. and {von Feilitzsch}, F. and {Weinz}, S. and {Wojcik}, M. and {Wurm}, M. and {Zaimidoroga}, O. and {Zavatarelli}, S. and {Zuber}, K. and {Zuzel}, G. and {Borexino Collaboration}},
	title = "{Simultaneous precision spectroscopy of p p , $^{7}$Be, and p e p solar neutrinos with Borexino Phase-II}",
	journal = {\prd},
	year = 2019,
	month = oct,
	volume = {100},
	number = {8},
	eid = {082004},
	pages = {082004},
	url = {https://doi.org/10.1103/PhysRevD.100.082004},
	doi = {10.1103/PhysRevD.100.082004},
	archivePrefix = {arXiv},
	eprint = {1707.09279},
	primaryClass = {hep-ex},
	adsurl = {https://ui.adsabs.harvard.edu/abs/2019PhRvD.100h2004A}
}

@ARTICLE{2024NuPhB100316473B,
	author = {{Baudis}, Laura},
	title = "{DARWIN/XLZD: A future xenon observatory for dark matter and other rare interactions}",
	journal = {Nuclear Physics B},
	year = 2024,
	month = jun,
	volume = {1003},
	eid = {116473},
	pages = {116473},
	url = {https://doi.org/10.1016/j.nuclphysb.2024.116473},
	doi = {10.1016/j.nuclphysb.2024.116473},
	archivePrefix = {arXiv},
	eprint = {2404.19524},
	primaryClass = {astro-ph.IM},
	adsurl = {https://ui.adsabs.harvard.edu/abs/2024NuPhB100316473B}
}

@ARTICLE{2021ScPP...10...30B,
	author = {{Bauer}, Martin and {Foldenauer}, Patrick and {Reimitz}, Peter and {Plehn}, Tilman},
	title = "{Light dark matter annihilation and scattering in LHC detectors}",
	journal = {SciPost Physics},
	year = 2021,
	month = feb,
	volume = {10},
	number = {2},
	eid = {030},
	pages = {030},
	url = {https://doi.org/10.21468/SciPostPhys.10.2.030},
	doi = {10.21468/SciPostPhys.10.2.030},
	archivePrefix = {arXiv},
	eprint = {2005.13551},
	primaryClass = {hep-ph},
	adsurl = {https://ui.adsabs.harvard.edu/abs/2021ScPP...10...30B}
}

@ARTICLE{2005ApJ...621L..85B,
	author = {{Bahcall}, John N. and {Serenelli}, Aldo M. and {Basu}, Sarbani},
	title = "{New Solar Opacities, Abundances, Helioseismology, and Neutrino Fluxes}",
	journal = {\apjl},
	year = 2005,
	month = mar,
	volume = {621},
	number = {1},
	pages = {L85-L88},
	url = {https://doi.org/10.1086/428929},
	doi = {10.1086/428929},
	archivePrefix = {arXiv},
	eprint = {astro-ph/0412440},
	primaryClass = {astro-ph},
	adsurl = {https://ui.adsabs.harvard.edu/abs/2005ApJ...621L..85B}
}

@ARTICLE{1987RvMP...59..671B,
	author = {{Bilenky}, S.~M. and {Petcov}, S.~T.},
	title = "{Massive neutrinos and neutrino oscillations}",
	journal = {Reviews of Modern Physics},
	year = 1987,
	month = jul,
	volume = {59},
	number = {3},
	pages = {671-754},
	url = {https://doi.org/10.1103/RevModPhys.59.671},
	doi = {10.1103/RevModPhys.59.671},
	adsurl = {https://ui.adsabs.harvard.edu/abs/1987RvMP...59..671B}
}

\end{document}